\documentclass[12pt]{article}
\pdfoutput=1
\usepackage[DIV13]{typearea}
\usepackage{indentfirst}
\RequirePackage{amsmath}
\RequirePackage{amssymb}
\RequirePackage{mathrsfs}
\RequirePackage{epsfig}
\RequirePackage{graphicx}
\usepackage{subfigure}
\usepackage{multicol}
\usepackage[usenames,dvipsnames]{color}
\usepackage{cite}
\RequirePackage[colorlinks=true
,urlcolor=blue
,anchorcolor=blue
,citecolor=blue
,filecolor=blue
,linkcolor=blue
,menucolor=blue
,pagecolor=blue
,linktocpage=true
,pdfproducer=medialab
]{hyperref}
\usepackage{cancel}
\newcommand{\email}[1]{\href{mailto:#1}{\tt #1}}
\numberwithin{equation}{section}
 \def\cO{{\cal O}}

\newcommand{\be}{\beta}

\def\be{\begin{equation}}
\def\ee{\end{equation}}
\def\beq{\begin{equation}}
\def\eeq{\end{equation}}
\def\bc{\begin{center}}
\def\ec{\end{center}}
\def\bea{\begin{eqnarray}}
\def\eea{\end{eqnarray}}
\def\dd{\displaystyle}
\def\nn{\nonumber}

\newcommand{\GeV}{\;\text{GeV}}

\begin{document}
\begin{titlepage}
\vspace*{-1cm}
\phantom{hep-ph/***} 
\flushright
\hfil{CERN-PH-TH/2012-137}
\hfil{RM3-TH/12-7}
\hfil{DFPD-2012/TH/4}
\hfil{TUM-HEP-833/12}\\

\vskip 1.5cm
\begin{center}
\mathversion{bold}
{\LARGE\bf Discrete Flavour Groups, $\theta_{13}$ and}\\[3mm]
{\LARGE\bf Lepton Flavour Violation}\\[3mm]
\mathversion{normal}
\vskip .3cm
\end{center}
\vskip 0.5  cm
\begin{center}
{\large Guido Altarelli}~$^{a,b)}$,
{\large Ferruccio Feruglio}~$^{c)}$,\\[2mm]
{\large Luca Merlo}~$^{d,e)}$,
{\large and Emmanuel Stamou}~$^{d,e,f)}$
\\
\vskip .7cm
{\footnotesize
$^{a)}$~Dipartimento di Fisica `E.~Amaldi', Universit\`a di Roma Tre, 
\\
INFN, Sezione di Roma Tre, I-00146 Rome, Italy
\\
\vskip .1cm
$^{b)}$~CERN, Department of Physics, Theory Division,
\\
CH-1211 Geneva 23, Switzerland
\\
\vskip .1cm
$^{c)}$~Dipartimento di Fisica `G.~Galilei', Universit\`a di Padova
\\
INFN, Sezione di Padova, Via Marzolo~8, I-35131 Padua, Italy
\vskip .1cm
$^{d)}$~Physik-Department, Technische Universit\"at M\"unchen, 
\\
James-Franck-Strasse, D-85748 Garching, Germany
\\
\vskip .1cm
$^{e)}$~
TUM Institute for Advanced Study, Technische Universit\"at M\"unchen, \\
Lichtenbergstrasse 2a, D-85748 Garching, Germany
\vskip .1cm
$^{f)}$~~Excellence Cluster Universe, Technische Universit\"at M\"unchen,\\
Boltzmannstrasse 2, D-85748 Garching, Germany
\vskip .5cm
\begin{minipage}[l]{.9\textwidth}
\begin{center} 
\textit{E-mail:} 
\email{guido.altarelli@cern.ch}, \email{feruglio@pd.infn.it},\\ \qquad\qquad\quad
\email{luca.merlo@ph.tum.de}, \email{emmanuel.stamou@ph.tum.de}
\end{center}
\end{minipage}
}
\end{center}
\vskip 1cm
\begin{abstract}
Discrete flavour groups have been studied in connection with special patterns of neutrino mixing suggested by the data, such as Tri-Bimaximal mixing (groups $A_4$, $S_4$\ldots) or Bi-Maximal mixing (group $S_4$\ldots) etc. We review the predictions for $\sin{\theta_{13}}$ in a number of these models and confront them with the experimental measurements. We compare the performances of the different classes of models in this respect. We then consider, in a supersymmetric framework, the important implications of these flavour symmetries  on lepton flavour violating processes, like $\mu \rightarrow e \gamma$ and similar processes. We discuss how the existing limits constrain these models, once their parameters are adjusted so as to optimize the agreement with the measured values of the mixing angles. In the simplified CMSSM context, adopted here just for indicative purposes, the small $\tan{\beta}$ range and heavy SUSY mass scales are favoured by lepton flavour violating processes, which makes it even more difficult to reproduce the reported muon $g-2$ discrepancy. 
\end{abstract}
\end{titlepage}
\setcounter{footnote}{0}

\pdfbookmark[1]{Table of Contents}{tableofcontents}
\tableofcontents

%
%%%%%%%%%%%%%%%%%%%%%%%%%   1.  Introduction       %%%%%%%%%%%%%%%%%%%%%%%%
%
\section{Introduction}

Neutrino mixing \cite{Altarelli:2004za,Mohapatra:2006gs,Grimus:2006nb,GonzalezGarcia:2007ib,Altarelli:2009wt,Altarelli:2010fk} is important because it could in principle provide new clues for the understanding of the flavour problem. Even more so since the neutrino mixing angles show a pattern that is completely different than that of quark mixing. The bulk of the data on neutrino oscillations are well described in terms of three active neutrinos. By now all three mixing angles have been measured,  although with different levels of accuracy (see Table~\ref{tab:data} \cite{Fogli:2012ua}; see also Ref.~\cite{Tortola:2012te}). In particular, we now have firm experimental evidence for a non-vanishing value of the smallest angle $\theta_{13}$ and a rather precise determination of its range (see Table~\ref{tab:t13} \cite{Abe:2011sj,Adamson:2011qu,Abe:2011fz,An:2012eh}).

\begin{table}[h]
\begin{center}
\begin{tabular}{|c|c|}
   \hline
  &\\[-2mm]
  $\Delta m^2_{sun}~(10^{-5}~{\rm eV}^2)$ 			&$7.54^{+0.26}_{-0.22}$   \\[1mm]
  $\Delta m^2_{atm}~(10^{-3}~{\rm eV}^2)$ 			&$2.43^{+0.06}_{-0.10}$ ($2.42^{+0.07}_{-0.11}$)   \\[1mm]
  $\sin^2\theta_{12}$ 											&$0.307^{+0.018}_{-0.016}$  \\[1mm]
  $\sin^2\theta_{23}$ 											&$0.386^{+0.024}_{-0.021}$ ($0.392^{+0.039}_{-0.022}$) \\[1mm]
  $\sin^2\theta_{13}$ 											&$0.0241^{+0.0025}_{-0.0025}$ ($0.0244^{+0.0023}_{-0.0025}$) 	  \\[1mm]
  $\delta_{CP}/\pi$ 												&$1.08^{+0.28}_{-0.31}$ ($1.09^{+0.38}_{-0.26}$)	  \\[1mm]
  \hline
  \end{tabular}
\caption{\label{tab:data} Fits to neutrino oscillation data from Ref.~\cite{Fogli:2012ua}. The results for both the normal and the inverse (in the brackets) hierarchies are shown.}
\end{center}
\end{table}

\begin{table}[h!]
\begin{center}
\begin{tabular}{|c|c|c|}
  \hline
  &&\\[-3mm]
  Quantity 	& $\sin^22\theta_{13}$ & $\sin^2\theta_{13}$ \\[1mm]
  \hline
  &&\\[-2mm]
  T2K\cite{Abe:2011sj} 					& 	$0.11^{+0.11}_{-0.05}$ ($0.14^{+0.12}_{-0.06}$)				
  													& 	$0.028^{+0.019}_{-0.024}$ ($0.036^{+0.022}_{-0.030}$) \\[1mm]
  MINOS\cite{Adamson:2011qu} 	&	$0.041^{+0.047}_{-0.031}$ ($0.079^{+0.071}_{-0.053}$)		
  													& 	$0.010^{+0.012}_{-0.008}$ ($0.020^{+0.019}_{-0.014}$) \\[1mm]
  DC\cite{Abe:2011fz} 					&	$0.086\pm0.041\pm0.030$ 				& $0.022^{+0.019}_{-0.018}$ 			\\[1mm]
  DYB\cite{An:2012eh}					&	$0.092\pm0.016\pm0.005$				& $0.024\pm0.005	$ 							\\[1mm]
  RENO\cite{Ahn:2012nd}				&	$0.113\pm0.013\pm0.019$				& $0.029\pm0.006	$ 			\\[1mm]
  \hline
  \end{tabular}
\caption{\label{tab:t13} The reactor angle measurements from the recent experiments T2K\cite{Abe:2011sj}, MINOS\cite{Adamson:2011qu}, DOUBLE CHOOZ\cite{Abe:2011fz}, Daya Bay \cite{An:2012eh} and RENO \cite{Ahn:2012nd}, for the normal (inverse) hierarchy.}
\end{center}
\end{table}

\begin{figure}[h!]
\centering
\hspace{-4mm}
\includegraphics[width=10.0 cm]{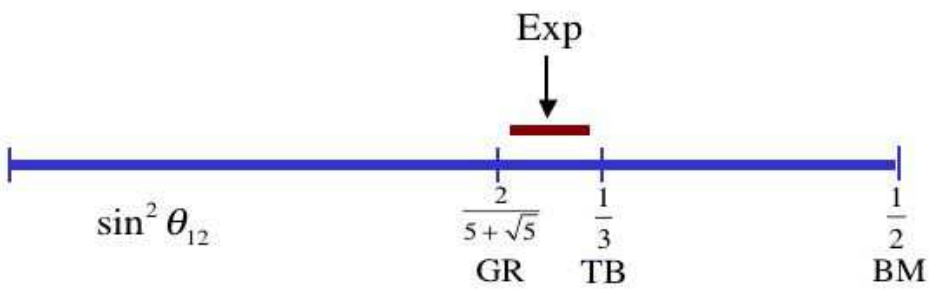}
\caption{The values of $\sin^2{\theta_{12}}$ for TB or GR or BM mixing are compared with the data at $1\sigma$.}
\label{fig:theta12}
\end{figure}

Models of neutrino mixing based on discrete flavour groups have received a lot of attention in recent years \cite{Altarelli:2010gt,Ishimori:2010au,Ludl:2010bj,Grimus:2010ak,Parattu:2010cy,Grimus:2011fk}. There are a number of special mixing patterns that have been studied in that context. The corresponding mixing matrices all have $\sin^2{\theta_{23}}=1/2$, $\sin^2{\theta_{13}}=0$, values that are good approximations to the data, and differ by the value of the solar angle $\sin^2{\theta_{12}}$.  The observed $\sin^2{\theta_{12}}$, the best measured mixing angle,  is very close, from below, to the so called Tri-Bimaximal (TB) value \cite{Harrison:2002er,Harrison:2002kp,Xing:2002sw,Harrison:2002et,Harrison:2003aw} of $\sin^2{\theta_{12}}=1/3$   (see Fig.~\ref{fig:theta12}). Alternatively, it is also very close, from above, to the Golden Ratio (GR) value \cite{Kajiyama:2007gx,Everett:2008et,Ding:2011cm,Feruglio:2011qq} $\sin^2{\theta_{12}}=\frac{1}{\sqrt{5}\,\phi} = \frac{2}{5+\sqrt{5}}\sim 0.276$, where $\phi= (1+\sqrt{5})/2$ is the GR (for a different connection to the GR, see Refs.~\cite{Rodejohann:2008ir,Adulpravitchai:2009bg}). On a different perspective, one has also considered models with Bi-Maximal (BM) mixing, where $\sin^2{\theta_{12}}=1/2$, i.e. also maximal, as the neutrino mixing matrix before diagonalization of charged leptons. This is in line with the well-known empirical observation that $\theta_{12}+\theta_C\sim \pi/4$, where $\theta_C$ is the Cabibbo angle, a relation known as quark-lepton complementarity \cite{Altarelli:2004jb,Raidal:2004iw,Minakata:2004xt,Frampton:2004vw,Ferrandis:2004vp,Kang:2005as,Li:2005ir,Cheung:2005gq,Xing:2005ur,Datta:2005ci,Antusch:2005ca,Lindner:2005pk,Minakata:2005rf,Ohlsson:2005js,King:2005bj,Dighe:2006zk,Chauhan:2006im,Hochmuth:2006xn,Schmidt:2006rb,Plentinger:2006nb,Plentinger:2007px}. Probably the exact complementarity relation becomes more plausible if replaced by $\theta_{12}+\mathcal{O}(\theta_C)\sim \pi/4$ (which we call ``weak'' complementarity). One can think of models where a suitable symmetry enforces BM mixing in the neutrino sector at leading order (LO) and the necessary, rather large, corrective terms to $\theta_{12}$ arise from the diagonalization of the charged lepton mass matrices \cite{Raidal:2004iw,Minakata:2004xt,Minakata:2005rf,Frampton:2004vw,Ferrandis:2004vp,Kang:2005as,Altarelli:2004jb,Li:2005ir,Cheung:2005gq,Xing:2005ur,Datta:2005ci,Ohlsson:2005js,Antusch:2005ca,Lindner:2005pk,King:2005bj,Dighe:2006zk,Chauhan:2006im,Schmidt:2006rb,Hochmuth:2006xn,Plentinger:2006nb,Plentinger:2007px,Altarelli:2009gn,Toorop:2010yh,Patel:2010hr,Meloni:2011fx,Shimizu:2010pg,Ahn:2011yj}. Thus, if one or the other of these coincidences is taken seriously, models where TB or GR or BM mixing is naturally predicted are a good first approximation. 

In the following we will mainly refer to TB or BM mixing which are the most studied first approximations to the data. The simplest symmetry that, at LO, leads to TB is $A_4$ while BM can be obtained from $S_4$.
$A_4$ models  have been studied widely (for a review and a list of references, see Ref.~\cite{Altarelli:2010gt}). At LO the typical $A_4$ model leads to exact TB mixing.  The LO approximation is then corrected by non-leading effects. Given the set of flavour symmetries and having specified the field content, the non-leading corrections to TB mixing, arising from higher dimensional effective operators, can be evaluated in a well-defined expansion. In the absence of specific dynamical tricks, in a generic model all three mixing angles receive corrections of the same order of magnitude. Since the experimentally allowed departures of
 $\theta_{12}$ from the TB value, $\sin^2{\theta_{12}}=1/3$, are small, numerically not larger than $\mathcal{O}(\lambda_C^2)$ where $\lambda_C=\sin\theta_C$, it follows that both $\theta_{13}$ and the deviation of $\theta_{23}$ from the maximal value are also expected to be typically of the same general size. The same qualitative conclusion also applies to $A_5$ models with GR mixing. This generic prediction of a small $\theta_{13}$, numerically of  $\mathcal{O}(\lambda_C^2)$, is now confronted with the most recent data. The central value $\sin{\theta_{13}} \sim 0.15$, from Table~\ref{tab:data}, is between $\mathcal{O}(\lambda_C^2) \sim \mathcal{O}(0.05)$ and $\mathcal{O}(\lambda_C) \sim \mathcal{O}(0.23)$. Since $\lambda_C$ is not that small, this gap is not too large and one can argue that models based on TB (or GR) mixing are still viable with preference for the lower side of the experimental range.

Of course, one can introduce additional theoretical input to improve the value of $\theta_{13}$ (for an updated list of recent models of this kind see Ref.~\cite{Varzielas:2012ss} and references therein; for recent models in the specific context of holographic composite Higgs see Refs.~\cite{Hagedorn:2011un,Hagedorn:2011pw}). In the case of $A_4$, one particularly interesting example is provided by the  Lin model \cite{Lin:2009bw} (see also Ref.~\cite{Varzielas:2010mp}), formulated before the T2K, MINOS, DOUBLE CHOOZ, Daya Bay and RENO results.  In the Lin model the $A_4$ symmetry breaking is arranged, by suitable additional $Z_n$ parities, in a way that the corrections to the charged lepton and the neutrino sectors are kept separated not only at LO but also at next-to-leading order (NLO). This way, the contribution to neutrino mixing from the diagonalization of the charged leptons can be of $\mathcal{O}(\lambda_C^2)$, while those in the neutrino sector of $\mathcal{O}(\lambda_C)$. In addition, in the Lin model these large corrections do not affect $\theta_{12}$ and satisfy the relation $\sin^2{\theta_{23}} =1/2 +1/\sqrt{2}\cos{\delta} \sin{\theta_{13}}$, with $\delta$ being the CKM-like CP violating phase of the lepton sector. Thus, in the Lin model the NLO corrections to the solar angle $\theta_{12}$ and to the reactor angle $\theta_{13}$ can be naturally of different orders. 

Alternatively, one can think of models where, because of a suitable symmetry,  BM mixing holds in the neutrino sector at LO and the corrective terms for $\theta_{12}$, which in this case are necessarily rather large, arise from the diagonalization of charged lepton masses \cite{Raidal:2004iw,Minakata:2004xt,Minakata:2005rf,Frampton:2004vw,Ferrandis:2004vp,Kang:2005as,Altarelli:2004jb,Li:2005ir,Cheung:2005gq,Xing:2005ur,Datta:2005ci,Ohlsson:2005js,Antusch:2005ca,Lindner:2005pk,King:2005bj,Dighe:2006zk,Chauhan:2006im,Schmidt:2006rb,Hochmuth:2006xn,Plentinger:2006nb,Plentinger:2007px,Altarelli:2009gn,Toorop:2010yh,Patel:2010hr,Meloni:2011fx,Shimizu:2010pg,Ahn:2011yj}. These terms from the charged lepton sector, numerically of order $\mathcal{O}(\lambda_C)$, would then generically also affect $\theta_{13}$. The resulting value could well be compatible with the present experimental values of $\theta_{13}$. An explicit model of this type based on the group $S_4$ has been developed in Ref.~\cite{Altarelli:2009gn} (see also Refs.~\cite{Toorop:2010yh,Patel:2010hr,Meloni:2011fx}). An important feature of this model is that only $\theta_{12}$ and $\theta_{13}$ are corrected by terms of $\mathcal{O}(\lambda_C)$ while $\theta_{23}$ is unchanged at this order. This model is compatible with present data and clearly prefers the upper range of the present experimental result for $\theta_{13}$.

In this work we discuss three possible classes of models: 1) typical $A_4$ models where $\theta_{13}$ is generically expected to be small, of the order of the observed departures of $\theta_{12}$ from the TB value, and thus with preference for the lower end of the allowed experimental range. 2) special $A_4$ models, like the Lin model, where $\theta_{13}$ is disentangled from the deviation of $\sin^2{\theta_{12}}$ from the TB value $1/3$ and can be as large as the upper end of the allowed experimental range. We give a general characterization of these special $A_4$ models where the dominant corrections to TB mixing do not arise from the charged lepton sector but from the neutrino sector. 3) Models where BM mixing holds in the neutrino sector and large corrections to $\theta_{12}$ and $\theta_{13}$ arise from the diagonalization of charged leptons. The value of $\theta_{13}$ could naturally be close to the present experimental range. In each of these possible models the dominant corrections to the LO mixing pattern involve a number of parameters of the same order of magnitude,  $\xi$. We discuss the success rate corresponding to the optimal value of $\xi$ for each model, obtained by scanning the parameter space according to a similar procedure for all three cases. We argue that, while the absolute values of the success rates depend on the scanning assumptions, their relative values in the three classes of models, provide a reliable criterium for comparison. We find that, for reproducing the mixing angles, the Lin type models have the best performance, as expected, followed by the typical $A_4$  models while the BM mixing models lead to an inferior score, as they most often fail to reproduce $\theta_{12}$.

We then discuss the implications for lepton flavour-violating (LFV) processes of the above three classes of possibilities, assuming a supersymmetric context, with or without See-Saw. The  present bounds pose severe constraints on the parameter space of the models (for a recent general analysis on model-independent flavour violating effects in the context of flavour models, see Ref.~\cite{Calibbi:2012at}). In particular, we refer to the recent improved MEG result \cite{Adam:2011ch} on the $\mu \rightarrow e \gamma$ branching ratio, $Br(\mu \rightarrow e \gamma) \lesssim 2.4\times10^{-12}$ at $95\%$ C.L. and to other similar processes like $\tau \rightarrow (e~\rm{or}~ \mu)  \gamma$. One expects that lepton flavour-violating processes may also have a large discriminating power in assessing the relative merits of the different models. We have studied this by adopting the simple CMSSM framework. While this overconstrained version of supersymmetry is rather marginal after the results of the LHC searches, more so if the Higgs mass really is around $m_H=125$ GeV, we still believe it can be used here for our indicative purposes. We find that the most constrained versions are the models with BM mixing at LO where relatively large corrections directly appear in the off-diagonal terms of the charged lepton mass matrix. The $A_4$ models turn out to be the best suited to satisfy the experimental bounds, as the non-diagonal charged lepton matrix elements needed to reproduce the mixing angles are quite smaller. An intermediate score is achieved by the models of the Lin type, where the main corrections to the mixing angles arise from the neutrino sector. Overall the $A_4$ models emerge well from our analysis and in particular those of the Lin type perhaps appear as the most realistic approach to the data among the discrete flavour group models that we have studied.  As for the regions of the CMSSM parameter space that are indicated by our analysis, the preference is for small $\tan{\beta}$ and large SUSY masses (at least one out of $m_0$ and $m_{1/2}$ must be above 1 TeV).  As a consequence it appears impossible, at least within the CMSSM model, to satisfy the MEG bound and simultaneously reproduce the muon $g-2$ discrepancy.

The paper is organized as follows. In Sect.~\ref{sec:Models} we discuss the models and their predictions for $\theta_{13}$ to then compare them with the data. In Sect.~\ref{sec:LFVformalism} we list the effective operators that induce lepton flavour violation and derive their contributions to the measured quantities. We then apply the general formalism to the specific models and observables. In Sect.~\ref{sec:Conclusions} we derive our conclusions.

%%%%%%%%%%%%%%%%%%%%%%%%%%%%%%%%%%%%%%%%%%%%%%%%%%%%%%%%%%%%%%%%%
%%%%%%%%%%%%%%%%%%%%%%%%%   2.  Models       
%%%%%%%%%%%%%%%%%%%%%%%%%%%%%%%%%%%%%%%%%%%%%%%%%%%%%%%%%%%%%%%%%
\section{Models}
\label{sec:Models}

We consider models invariant under a flavour symmetry group $G_f$.
At the LO the lepton mixing arises from a mismatch between the residual symmetries $G_e$ and $G_\nu$ of charged lepton and neutrino sectors, respectively.
In this approximation charged leptons and neutrinos acquire mass from two independent sets of flavons, $\Phi_e$ and $\Phi_\nu$, whose VEVs preserve two Abelian groups: $G_e=Z_n$ $(n\ge 3)$ and $G_\nu=Z_2\times Z_2$. The groups $G_e$ and $G_\nu$ can be subgroups of $G_f$ or, as a result of the specific field content of the model and of the LO approximation,
they can contain some accidental symmetry and generate a group $G$ different from (and possibly larger than) $G_f$.\footnote{
In the present paper we concentrate only on the TB, GR or BM patterns, originated by the mismatch between $G_e$ and $G_\nu$. However, interesting deformations of these patterns, all predicting already at the LO a non-vanishing $\theta_{13}$, arise considering the finite modular groups $\Gamma_N$, $N>1$: in Refs.~\cite{Toorop:2011jn,deAdelhartToorop:2011re}, a comprehensive analysis is presented for an arbitrary $G_e$ and $G_\nu = Z_2 \times Z_2$. In Refs.~\cite{Ge:2011ih,Ge:2011qn,Hernandez:2012ra}, a more general study has been presented, where the residual symmetry in the neutrino sector is $G_\nu=Z_2$, while the other $Z_2$ component arises accidentally.} Lepton mass matrices can be expanded in inverse powers of the cut-off scale $\Lambda$
\beq
\begin{aligned}
m_e&=m_e^{(0)}+\delta m_e^{(1)}+\ldots\\
m_\nu&=m_\nu^{(0)}+\delta m_\nu^{(1)}+\ldots
\end{aligned}
\label{masses}
\eeq
where $\delta m_{e,\nu}^{(1)}/m_{e,\nu}^{(0)}=\cO(\langle\Phi_{e,\nu}\rangle/\Lambda)$ and dots stand for higher-order terms. The LO contributions $m_e^{(0)}$and $m_\nu^{(0)}$ depend only on $\langle \Phi_e\rangle$ and $\langle\Phi_\nu\rangle$, respectively.
They are invariant under $G_e$ and $G_\nu$:
\beq
\begin{aligned}
\rho(g_{e i})^\dagger m^{(0) \dagger}_e m_e^{(0)} \rho(g_{e i})&=m^{(0)\dagger}_e m_e^{(0)}\,,\\
\rho(g_{\nu i})^T m_\nu^{(0)}~ \rho(g_{\nu i})&=m_\nu^{(0)}\,.
\end{aligned}
\label{AB}
\eeq
Here $g_{ei}$ and $g_{\nu i}$ are the elements of $G_e$ and $G_\nu$ and $\rho$ denotes an irreducible three-dimensional unitary representation of the group $G$ generated by $G_e$ and $G_\nu$.\footnote{Later on we will include in $m_e^{(0)}$ also higher-order contributions satisfying the property (\ref{AB}).}
Both $G_e$ and $G_\nu$ are Abelian and the matrices $\rho(g_{e i})$ and $\rho(g_{\nu i})$ can be diagonalized by two independent unitary transformations $\Omega_e$ and $\Omega_\nu$ 
\be
\rho(g_{\nu i})_{diag}=\Omega_\nu^\dagger~ \rho(g_{\nu i})~ \Omega_\nu\,,\qquad\qquad
\rho(g_{e i})_{diag}=\Omega_e^\dagger~ \rho(g_{e i})~ \Omega_e\,,
\ee
and the mixing matrix $U_{PMNS}$ reflects the misalignment between the two bases: 
\be
U_{PMNS}=\Omega_e^\dagger \Omega_\nu\,.
\label{LOmix}
\ee
It is immediate that the mixing matrix is independent of the choice of the basis. We choose to work in the basis where $\rho(g_{ei})$ and $m^{(0)\dagger}_e m_e^{(0)}$
are diagonal: $\Omega_e=1$.
Beyond LO, the prediction for the mixing in eq.~(\ref{LOmix}) is modified. In general the VEVs of $\Phi_e$ and $\Phi_\nu$ are corrected by terms of relative order $\langle\Phi_{e,\nu}\rangle/\Lambda$ and do not preserve $G_e$ and $G_\nu$ any more.
Moreover higher order operators contribute to lepton masses without respecting the LO residual symmetries. The NLO corrections are suppressed with respect to the LO contributions by the ratio between the flavon VEVs ($\langle \Phi_e\rangle,\langle\Phi_\nu\rangle$) and $\Lambda$. Depending on the agreement of the LO approximation to the data, $\langle \Phi_e\rangle/\Lambda$ and $\langle\Phi_\nu\rangle/\Lambda$ will typically range between $\lambda_C^2$ and $\lambda_C$.

%%%%%%%%%%%%%%%%%%%%%%%%%   2.1  $A_4$ models      
\boldmath
\subsection{$A_4$ Models}
\unboldmath

We refer to SUSY models based on the flavour symmetry $G_f=A_4\times G_{AUX}$, where the $G_{AUX}$ factor depends on the specific realization \cite{Lin:2009bw,Altarelli:2005yp,Altarelli:2005yx,Altarelli:2009kr}.\footnote{The present analysis applies also to models based on flavour symmetries that contain the group $A_4$, such as $S_4$: few examples can be found in Refs.~\cite{Bazzocchi:2009pv,Bazzocchi:2009da,Ding:2009iy,Meloni:2009cz}, where the TB pattern is predicted at the LO, while at the NLO the mixing is corrected in a similar way as we are going to discuss in this section. Moreover, our analysis applies also to the model based on the flavour group $T'$, described in Ref.~\cite{Feruglio:2007uu}, even if $T'$ does not contain $A_4$ as a subgroup. Although we could generalize our analysis to a brother class of symmetries, we focus on the flavour group $A_4$ that represents the minimal choice in terms of dimensions of a group.}
The group $A_4$ can be generated by two elements $S$ and $T$ satisfying
\be 
S^2=(ST)^3=T^3=1\,.
\ee
The irreducible representations of $A_4$ are a triplet $3$ and three inequivalent singlets $1$, $1'$ and $1''$.
In the triplet representation $S$ and $T$ can be chosen as:
\be
T=\left(
\begin{array}{ccc}
1&0&0\\
0&\omega^2&0\\
0&0&\omega
\end{array}
\right)\,,\qquad\qquad
S=\frac{1}{3}\left(
\begin{array}{ccc}
-1&2&2\\
2&-1&2\\
2&2&-1
\end{array}
\right)\,,
\ee
where $\omega=e^{i 2\pi/3}$.
Under $A_4$ the electroweak SU(2) lepton doublets $l$ transform as a triplet, while the electroweak singlets
$e^c$, $\mu^c$ and $\tau^c$ and the electroweak Higgs doublets $H_{u,d}$ as singlets.
In the flavon sector both $\Phi_e$ and $\Phi_\nu$ always include a triplet, but they can also include additional singlets. At the LO and in the exact SUSY limit the VEVs of $\Phi_e$ and $\Phi_\nu$ are determined by two separate sets of equations and satisfy at LO:
\be
T '\langle\Phi_e\rangle = \langle\Phi_e\rangle\,,\qquad\qquad
S\langle\Phi_\nu\rangle = \langle\Phi_\nu\rangle\,.
\label{sandt}
\ee
The transformation $T'$ can coincide with the $T$ generator of $A_4$ \cite{Altarelli:2005yp,Altarelli:2005yx}, or can represent an accidental symmetry of the charged lepton Lagrangian still satisfying ${T'}^3=1$, as in the models of Refs.~\cite{Lin:2009bw,Altarelli:2009kr}, where $T'=\omega T$.
The charged lepton mass matrix $m_e$ is given by  
\be
m_e= m_e^{(0)}+\delta m_e^{(1)}+\ldots
\ee
where
\be
m_e^{(0)}=v_d
\left(
\begin{array}{ccc}
y_e&0&0\\
0&y_\mu&0\\
0&0&y_\tau
\end{array}
\right)\eta\,.
\label{mezero}
\ee
Here $v_d$ is the VEV of $H_d$, $y_f$ $(f=e,\mu,\tau)$ are dimensionless quantities and $\eta$ is a small parameter that breaks the flavour symmetry $A_4$. At the LO the charged lepton mass matrix $m^{(0)}_e$, only depending on $\langle \Phi_e\rangle$, is diagonal and invariant under the action of the transformation $T'$:
\be
{T'}^\dagger m^{(0) \dagger}_e m_e^{(0)} T'=m^{(0)\dagger}_e m_e^{(0)}\,.
\ee
We have $G_e=Z_3$, generated by $T'$. The hierarchical pattern $y_e\ll y_\mu\ll y_\tau$ can be reproduced by requiring that operators of increasing dimension contribute to $y_\tau$, $y_\mu$ and $y_e$, respectively
\footnote{Here, we include in $m_e^{(0)}$ all contributions arising from the LO $\langle\Phi_e\rangle$, independently of the dimensionality of the corresponding operator.}.
This can be achieved either by means of a Froggatt-Nielsen $U(1)$ symmetry  contained in $G_{AUX}$ \cite{Altarelli:2005yp,Altarelli:2005yx} or through some discrete components of $G_{AUX}$ as in Refs.~\cite{Lin:2009bw,Altarelli:2009kr}.

This class of models can be realized both with and without a See-Saw mechanism. In the first case there are three right-handed neutrinos transforming as a triplet of $A_4$, while in the second case the source of neutrino masses
is a set of higher dimensional operators violating the total lepton number. In either case the light neutrino mass matrix $m_\nu$ is given by:
\be
m_\nu=m_\nu^{(0)}
+\delta m_\nu^{(1)}+\ldots
\ee
where
\be
m_\nu^{(0)}=\left(
\begin{array}{ccc}
x&y&y\\
y&x+z&y-z\\
y&y-z&x+z
\end{array}
\right)\,.
\label{mnulo}
\ee
The parameters $x$, $y$ and $z$ are quadratic in $v_u$, the VEV of $H_u$, and inversely proportional to the scale associated with the violation
of the total lepton number.
The LO term $m_\nu^{(0)}$ is invariant under the $Z_2\times Z_2$ symmetry generated by $S$, eq.~(\ref{sandt}), and by
\be
A_{23}=\left(
\begin{array}{ccc}
1&0&0\\
0&0&1\\
0&1&0
\end{array}
\right)\,.
\label{a23}
\ee
Indeed,
\be
S^T m_\nu^{(0)} S=m_\nu^{(0)}
\qquad\qquad\text{and}\qquad\qquad
A_{23}^T m_\nu^{(0)} A_{23}=m_\nu^{(0)}\,.
\ee
The $Z_2$ symmetry generated by the matrix $A_{23}$ is an accidental symmetry. 
The matrix $m_\nu^{(0)}$ of eq.~(\ref{mnulo}) is the most general one invariant under both $S$ and $A_{23}$. In the minimal formulation of Refs.~\cite{Lin:2009bw,Altarelli:2005yp,Altarelli:2005yx} the parameters $x$, $y$ and $z$ are not independent.
If neutrino masses are generated via the See-Saw mechanism \cite{Lin:2009bw,Altarelli:2005yx} they are related by
\be
z=\frac{(x-y)^2}{4y-x}\,.
\ee
If neutrino masses are parametrized directly through a higher dimensional operator \cite{Altarelli:2005yp,Altarelli:2005yx} we have
\be
z=-(x+2y)\,.
\ee 
At the LO $m_e^{(0)}$ is diagonal while $m_\nu^{(0)}$ is diagonalized by 
\be
U_{TB}=
\left(
\begin{array}{ccc}
\sqrt{\frac{2}{3}}&\sqrt{\frac{1}{3}}&0\\
-\sqrt{\frac{1}{6}}&\sqrt{\frac{1}{3}}&-\sqrt{\frac{1}{2}}\\
-\sqrt{\frac{1}{6}}&\sqrt{\frac{1}{3}}&\sqrt{\frac{1}{2}}
\end{array}
\right)\,.
\ee
This holds for any value of the parameters $x$, $y$ and $z$ since $U_{TB}$ simultaneously diagonalizes both $S$ and $A_{23}$.

The departure from the LO approximation depends on the subleading contributions $\delta m_e^{(1)}$ and $\delta m_\nu^{(1)}$, which can vary for different models.
In all models considered here \cite{Altarelli:2005yp,Altarelli:2005yx,Lin:2009bw,Altarelli:2009kr} the NLO correction to the charged lepton mass matrix is not invariant under $T'$ and is of the following type:
\be
\delta m_e^{(1)}=
v_d
\left(
\begin{array}{ccc}
\cO(y_e)&\cO(y_e)&\cO(y_e)\\
\cO(y_\mu)&\cO(y_\mu)&\cO(y_\mu)\\
\cO(y_\tau)&\cO(y_\tau)&\cO(y_\tau)
\end{array}
\right)\eta~\xi\,,
\ee
where $\xi$ is a small adimensional parameter given by the ratio between a flavon VEVs and $\Lambda$. The transformation to diagonalize $m_e$ is $V_e^T m_e U_e=m_e^{diag}$ with
\beq
U_e=
\left(
\begin{array}{ccc}
1							&c^e_{12}\,\xi			&c^e_{13}\,\xi\\
-c_{12}^{e*}\,\xi		&1							&c^e_{23}\,\xi\\
-c_{13}^{e*}\,\xi		&-c_{23}^{e*}\,\xi		&1\\
\end{array}
\right)
\label{ue}
\eeq
where $c^e_{12}$, $c^e_{13}$ and $c^e_{23}$ are complex parameters of order one in absolute value.
We discuss the NLO contribution to $m_\nu$ by distinguishing two cases.

%%%%%%%%%%%%%%%%%%%%%%%%%   2.1.1  Typical $A_4$ models  
\boldmath
\subsubsection{Typical $A_4$ Models}
\unboldmath
\label{sec:TypicalA4}

In ``typical'' $A_4$ models \cite{Altarelli:2005yx,Altarelli:2009kr}, the NLO contribution $\delta m_\nu^{(1)}$ is a generic symmetric matrix with entries suppressed, compared to the corresponding entries in $m_\nu^{(0)}$, by a relative factor $\xi'$,
of the order of the ratio between a flavon VEV and $\Lambda$. This occurs both with and without the See-Saw mechanism. The generic transformation that diagonalize $m_\nu$ is $U_{TB}\cdot U_\nu$ where
\beq
U_\nu=
\left(
\begin{array}{ccc}
1								&c^\nu_{12}\,\xi'				&c^\nu_{13}\,\xi'\\
-c_{12}^{\nu*}\,\xi'		&1									&c^\nu_{23}\,\xi'\\
-c_{13}^{\nu*}\,\xi'		&-c_{23}^{\nu*}\,\xi'			&1\\
\end{array}
\right)\,,
\label{unu}
\eeq
where $c^\nu_{12}$, $c^\nu_{13}$ and $c^\nu_{23}$ are complex parameters of order one in absolute value.
Barring a fine tuning of the Lagrangian parameters, in these models the suppression factors
$\xi$ and $\xi'$ are expected to be of the same order of magnitude. For example, beyond the LO the equations satisfied by $\langle\Phi_e\rangle$ and $\langle\Phi_\nu\rangle$ are no longer decoupled and the corrections
to the LO flavon VEVs are of the same size, for both $\Phi_e$ and $\Phi_\nu$. All the elements of the mixing matrix get corrections of the same size $\xi\approx \xi'$. We expect\footnote{Eq.~(\ref{sinNLOTB}) is a particular case of the general parametrization presented Ref.~\cite{King:2007pr}:
\beq
\sin\theta_{23}=\dfrac{1}{\sqrt2}(1+a)\,,\qquad
\sin\theta_{12}=\dfrac{1}{\sqrt3}(1+s)\,,\qquad
\sin\theta_{13}=\dfrac{r}{\sqrt2}\,,
\eeq
with $a$, $s$ and $r$ real numbers. The expressions in Eq.~(\ref{sinNLOTB}) show explicitly the dependence of the NLO mixing angles
on the corrections from both the neutrino and the charged lepton sectors.}:
\beq
\begin{aligned}
\sin^2\theta_{23}&=\frac{1}{2}+{\cal R}e(c^e_{23})\,\xi+\dfrac{1}{\sqrt{3}}\left({\cal R}e(c^\nu_{13})-\sqrt2\,{\cal R}e(c^\nu_{23})\right)\,\xi\\
\sin^2\theta_{12}&=\frac{1}{3}-\frac{2}{3}{\cal R}e(c^e_{12}+c^e_{13})\,\xi+\dfrac{2\sqrt2}{3}\,{\cal R}e(c^\nu_{12})\,\xi\\[1mm]
\sin\theta_{13}&=\dfrac{1}{6}\left|3\sqrt2\left(c^e_{12}-c^e_{13}\right)+2\sqrt3\left(\sqrt2\,c^\nu_{13}+c^\nu_{23}\right)\right|\,\xi\,.
\end{aligned}
\label{sinNLOTB}
\eeq

\begin{figure}[h!]
 \centering
   \includegraphics[width=9cm]{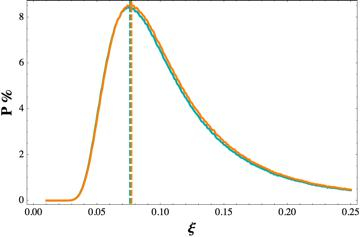}
 \caption{ {\bf Typical {\boldmath$A_4$\unboldmath} Models}. Success Rates as a function of the parameter $\xi$. The $c^{e,\nu}_{ij}$ parameters that multiply $\xi$ are treated as random complex numbers with absolute values following a Gaussian distribution around 1 with variance 0.5. In Cyan the NH and in Orange the IH.  The value of $\xi$ that maximizes the success rate is $0.076 (0.077)$ for NH (IH).}
\label{fig:SuccessRatesTB}
\end{figure}

We see that to reach the central value for the reactor angle in agreement with the value reported in Table~\ref{tab:data}, the parameter $\xi$ should be of $\cO(0.1)$. A precise value can be found studying the success rate to reproduce all the three mixing angles inside the corresponding $3\sigma$ ranges, depending on the value of $\xi$. As shown in Fig.~\ref{fig:SuccessRatesTB}, the value of $\xi$ that maximizes the success rate is $0.076 (0.077)$ for NH (IH). The corresponding value is $\sim 8.5 \%$, which is not large but not hopelessly small either.

\begin{figure}[h!]
 \centering
  \subfigure[Correlation between $\sin^2\theta_{12}$ and $\sin^2\theta_{13}$.]
   {\includegraphics[width=7.7cm]{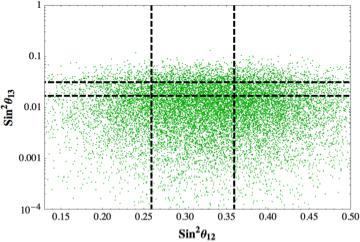}}
  \subfigure[Correlation between $\sin^2\theta_{23}$ and $\sin^2\theta_{13}$.]
   {\includegraphics[width=7.7cm]{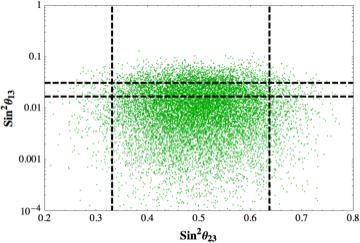}}
 \caption{ {\bf Typical {\boldmath$A_4$\unboldmath} Models.} On the left (right), we plot $\sin^2\theta_{13}$ as a function of $\sin^2\theta_{12}$ ($\sin^2\theta_{23}$), following eq.~(\ref{sinNLOTB}).  The dashed-black lines represent the $3\sigma$ values for the mixing angles from the Fogli {\it et al.} fit \cite{Fogli:2012ua}. Only the NH data sets is shown. The parameter $\xi$ is taken equal to $0.076$. The $c^{e,\nu}_{ij}$ parameters that multiply $\xi$ are treated as random complex numbers with absolute values following a Gaussian distribution around 1 with variance 0.5.}
 \label{fig:Sin12qvsSin13qTB_NIH}
\end{figure}

We analyze quantitatively the expressions in eq.~(\ref{sinNLOTB}) and their correlations in Fig.~\ref{fig:Sin12qvsSin13qTB_NIH}:  in the plots on the left (right), we show the correlation between $\sin^2\theta_{13}$ and $\sin^2\theta_{12}$ ($\sin^2\theta_{23}$). The parameter $\xi$ is taken equal to $0.076$. The $c^{e,\nu}_{ij}$ parameters that multiply $\xi$ are treated as random complex numbers with absolute values following a Gaussian distribution around 1 with variance 0.5. In the plots we show only the NH case. The IH case is similar.

As we can see, the plots are representing the general behaviour of this class of models: $\sin^2\theta_{13}$ increases with $\xi$, but correspondingly also the deviation of $\sin^2\theta_{12}$ from $1/3$ does. As a result, even for the value of $\xi$ that maximizes the success rate, the requirement for having a reactor angle inside its $3\sigma$ error range corresponds to a prediction for the solar angle that spans all the $3\sigma$ experimental error bar and is often not even in agreement with the data.

%%%%%%%%%%%%%%%%%%%%%%%%%   2.1.2  Special $A_4$ models  
\boldmath
\subsubsection{Special $A_4$ Models}
\label{sec:SpecialA4}
\unboldmath

In these models the accidental symmetry $A_{23}$ of the neutrino sector is broken by a relatively large amount so that, in first approximation, the residual symmetries 
of the charged lepton and neutrino sectors are those generated by $T'$ and $S$, respectively. At the LO and in the chosen basis $m_e^\dagger m_e$ is diagonal
while $m_\nu$ is invariant under $S$:
\be
S^T m_\nu S=m_\nu\,.
\ee
The most general solution to this constraint can be parametrized in the following form:
\be
m_\nu=
\left(
\begin{array}{ccc}
x&y-w&y+w\\
y-w&x+z+w&y-z\\
y+w&y-z&x+z-w
\end{array}
\right)\,,
\label{generalmnu}
\ee
We see that $w$ describes the deviation of $m_\nu$ from the form associated to the Tri-Bimaximal mixing, see eq. (\ref{mnulo}).
The matrix $m_\nu$ can be diagonalized in two steps. First we transform $m_\nu$ by a Tri-Bimaximal rotation:
\be
m_\nu'=U_{TB}^T m_\nu U_{TB}=
\left(
\begin{array}{ccc}
x-y&0 &\sqrt{3} w \\
0&x+2y &0 \\
\sqrt{3} w&0 &x-y+2z
\end{array}
\right)\,.
\ee
Second, we perform a unitary transformation in the (1,3) plane:
\be
V=\left(
\begin{array}{ccc}
\alpha & 0&\xi'\\
0&1&0\\
-\xi'^*&0&\alpha^*
\end{array}
\right)\,,\qquad\qquad
|\alpha|^2+|\xi'|^2=1\,,
\ee
\be
V^T m_\nu' V= m_\nu^{diag}\,.
\ee
The exact rotation is given by:
\be
\frac{2\alpha\xi'}{|\alpha|^2-|\xi'|^2}=\frac{u v^*(u^*-v)}{|v|^2-|u|^2}\,,\quad\text{with}\quad
u\equiv \frac{2\sqrt{3} w}{x-y}\,,\quad\text{and}\quad
v\equiv-\frac{2 \sqrt{3} w}{x-y+2z}\,.
\ee
The unitary matrix that diagonalizes $m_\nu$ is
\beq
U_{TB}\,V=
\left(
\begin{array}{ccc}
\sqrt{2/3}\,\alpha						&1/\sqrt{3} 		&\sqrt{2/3}\, \xi' \\
-\alpha/\sqrt{6}+\xi'^*/\sqrt{2}		&1/\sqrt{3} 		&-\alpha^*/\sqrt{2}-\xi'/\sqrt{6} \\
-\alpha/\sqrt{6}-\xi'^*/\sqrt{2}		&1/\sqrt{3} 		&+\alpha^*/\sqrt{2}-\xi'/\sqrt{6} \\
\end{array}
\right)\,.
\label{specialmixing}
\eeq
Such a mixing pattern is very interesting because the observed $\theta_{13}$ can be reproduced by choosing $\xi'$ of order 0.1 and the predicted value of $\sin^2\theta_{12}$
deviates from $1/3$ only by terms of order $\xi'^2$. To preserve these properties corrections from the charged lepton sector should be small compared to $\xi'$.
This can be realized by adopting two different expansion parameters $\xi\ll \xi'$ for the charged lepton sector and for the neutrino sector.
A model along these lines has been built in Ref. ~\cite{Lin:2009bw}. The setup is arranged in such a way that $\langle\Phi_e\rangle$ and $\langle\Phi_\nu\rangle$ satisfy decoupled equations up to NLO so that it is possible to achieve $\langle\Phi_e\rangle\ll \langle\Phi_\nu\rangle$
and to maintain the property of eq.~(\ref{sandt}) up to NLO. Moreover $\langle\Phi_\nu\rangle$ couples to charged leptons only at the NNLO so that the dominant source of 
corrections to the neutrino mixing pattern is $\delta m_\nu^{(1)}$, which in turns, being dominated by $\langle\Phi_\nu\rangle$, is invariant under $S$.

\begin{figure}[h!]
 \centering
 \includegraphics[width=9cm]{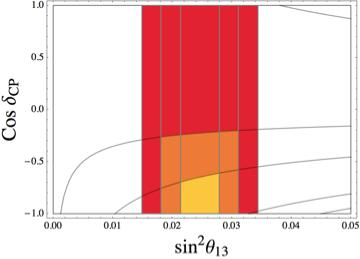}
 \caption{ {\bf Special {\boldmath$A_4$\unboldmath} Models.} Contour plot for $1,2,3\sigma$ values of $\sin^2\theta_{23}$ in the parameter space $\sin^2\theta_{13}$--$\cos\delta_{CP}$ according to the values for the NH in Table~\ref{tab:data}. The Yellow, Orange, Red regions refer to the data at $1\sigma$, $2\sigma$, $3\sigma$, respectively.}
 \label{fig:CorrelationsTBLin}
\end{figure}

In eq. (\ref{specialmixing}) it is not restrictive to choose $\alpha$ real and positive and we have:
\bea
\delta_{CP}&\approx&\arg\xi'\\
\sin\theta_{13}&=&\left|\sqrt{\frac{2}{3}}\,\xi'+\frac{c^e_{12}-c^e_{13}}{\sqrt{2}}\,\xi\right|\\
\sin^2\theta_{12}&=&\frac{1}{3-2\, |\xi'|^2}-\frac{2}{3}\,{\cal R}e(c^e_{12}+c^e_{13})\,\xi\nn\\
&=&\frac{1}{3}+\frac{2}{9}\,|\xi'|^2-\frac{2}{3}\,{\cal R}e(c^e_{12}+c^e_{13})\,\xi
\label{sinNNLOTBLinSol}\\
\sin^2\theta_{23}&=&\frac{1}{2}
\frac{\left(1+\frac{\xi'}{\sqrt{3}\,\alpha}\right)\left(1+\frac{\xi'^*}{\sqrt{3}\,\alpha}\right)}
{\left(1+\frac{|\xi'|^2}{3\,\alpha^2}\right)}+{\cal R}e(c^e_{23})\,\xi\nn\\
&=&\frac{1}{2}+\frac{1}{\sqrt{3}}\,|\xi'|\,\cos\delta_{CP}+{\cal R}e(c^e_{23})\,\xi
\label{sinNNLOTBLinAtm}
\eea
where we have also included the effects coming from the diagonalization of the charged lepton sector as in eq.~(\ref{ue}), to first order in $\xi$. 
The second equality shows the result expanded in powers of $|\xi'|$, to the order $|\xi'|^2$. In these models $|\xi'|$ is chosen to be of order $0.1$, larger than $\xi$ so that the contribution of eq.~(\ref{ue}) can be neglected, and the lepton mixing matrix is very close to $U_{TB}\,V$.
It is interesting to note that if we neglect the corrections proportional to $\xi$, we have exact correlations between the reactor angle and the other two angles\footnote{It has been shown in Ref.~\cite{Hernandez:2012ra}, from general group theoretical considerations, that these correlations are a general feature of flavour models when the symmetry group of the charged lepton (neutrino) mass matrix is $Z_3$ ($Z_2$).}:
\beq
\sin^2\theta_{12}=\frac{1}{3(1-\sin^2\theta_{13})}\,,\qquad\qquad
\sin^2\theta_{23}=\frac{1}{2}+\frac{1}{\sqrt2}\sin\theta_{13}\cos\delta_{CP}\,.
\eeq
The first expression shows that the unitary transformation $V$ always increases the solar angle from the TB value, while the preferred 1$\sigma$ interval is below the TB prediction, see Fig. \ref{fig:theta12}. This is a small effect, of second order in $\theta_{13}$, that can be compensated by the corrections proportional to $\xi$. The second correlation involves the Dirac CP phase and is particularly interesting considering the recent hint of a CP phase close to $\pi$ for the NH case. In Fig.~\ref{fig:CorrelationsTBLin}, we graphically compare this second expression with the present data for the NH case: when considering the $1\sigma$ ($2\sigma$) ranges for the mixing angles, one sees an indication that $\cos\delta_{CP}$ lies in the interval $[-1,-0.5]$, while no indication arises when the $3\sigma$ error band for $\sin^2\theta_{23}$ is taken into account. Although these results for the CP phase is modified by the inclusion of the subleading $\xi$ contributions, these correlations will allow an interesting test for such models once $\delta_{CP}$ is measured and the precision on $\sin^2\theta_{23}$ is improved.

\begin{figure}[h!]
 \centering
   \includegraphics[width=9cm]{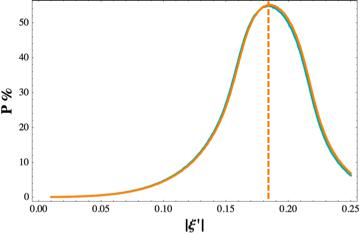}
 \caption{ {\bf Special {\boldmath$A_4$\unboldmath} Models.} Success Rates as a function of the parameter $|\xi'|$. The parameters have been chosen such that $\xi$ is a real number in $[0.005,\,0.06]$ and $c^e_{ij}$ are random complex numbers with absolute values following a Gaussian distribution around 1 with variance 0.5. In Cyan the NH and in Orange the IH. The value of $|\xi'|$ that maximizes the success rate for both NH and IH is $0.184$.}
 \label{fig:SuccessRatesTBLin}
\end{figure}

In Fig.~\ref{fig:SuccessRatesTBLin}, we study the success rate to reproduce all the three mixing angles inside their corresponding $3\sigma$ error ranges, as a function of $|\xi'|$. The parameters have been chosen such that $\xi$ is a real number in $[0.005,\,0.06]$ and $c^e_{ij}$ are random complex numbers with absolute values following a Gaussian distribution around 1 with variance 0.5. The value of $|\xi'|$ that maximizes the success rate for both NH and IH is $0.184$. The corresponding success rate is much larger in these models ($\sim 55 \%$) than for the typical $A_4$ models.

\begin{figure}[h!]
 \centering
   \subfigure[Correlation between $\sin^2\theta_{12}$ and $\sin^2\theta_{13}$.]
   {\includegraphics[width=7.7cm]{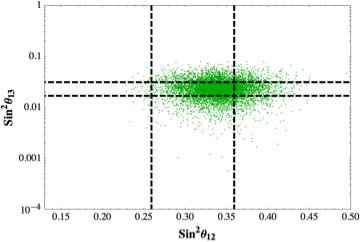}}
   \subfigure[Correlation between $\sin^2\theta_{23}$ and $\sin^2\theta_{13}$.]
   {\includegraphics[width=7.7cm]{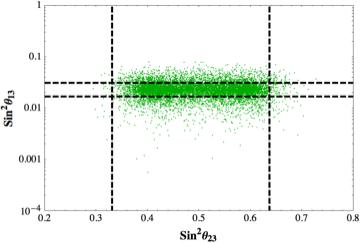}}
 \caption{ {\bf Special {\boldmath$A_4$\unboldmath} Models.} $\sin^2\theta_{13}$ as a function of $\sin^2\theta_{12}$ ($\sin^2\theta_{23}$) is plotted on the left (right), following eqs.~(\ref{sinNNLOTBLinSol}) and (\ref{sinNNLOTBLinAtm}). The dashed-black lines represent the $3\sigma$ values for the mixing angles from the Fogli {\it et al.} fit \cite{Fogli:2012ua}. Only the NH data sets is shown. The parameter $\xi$ is a real number in $[0.005,\,0.06]$; $\xi'$ is a complex number with absolute values equal to $0.184$; the parameters $c^e_{ij}$ are random complex numbers with absolute values following a Gaussian distribution around 1 with variance 0.5.}
 \label{fig:Sin23qvsSin13qTBLin_NIH}
\end{figure}

We analyze quantitatively the deviations in eqs.~(\ref{sinNNLOTBLinSol}) and (\ref{sinNNLOTBLinAtm}) and their correlations in Fig.~\ref{fig:Sin23qvsSin13qTBLin_NIH}: in the plots on the left (right) column, we show the correlations in eqs.~(\ref{sinNNLOTBLinSol}) and (\ref{sinNNLOTBLinAtm}) between $\sin^2\theta_{13}$ and $\sin^2\theta_{12}$ or $\sin^2\theta_{23}$, respectively. The parameters have been chosen such that $\xi$ is a real number in $[0.005,\,0.06]$; $\xi'$ is a complex number with absolute values equal to $0.184$; the parameters $c^e_{ij}$ are random complex numbers with absolute values following a Gaussian distribution around 1 with variance 0.5. In the plots we show only the NH case. The IH case is similar. For this choice of the parameters, the model can well describe all three angles inside the corresponding $3\sigma$ interval, and its success rate is much larger than that of the typical TB models, as turns out by comparing Figs.~\ref{fig:SuccessRatesTB} and \ref{fig:SuccessRatesTBLin}.

%%%%%%%%%%%%%%%%%%%%%%%%%   2.2  $S_4$ models      
\boldmath
\subsection{$S_4$ Models}
\unboldmath

In this section we refer to a SUSY model based on the flavour symmetry $G_f=S_4\times Z_4\times U(1)$ \cite{Altarelli:2009gn}, but we keep the presentation slightly more general, to embrace
a wider class of possibilities \cite{Toorop:2010yh,Patel:2010hr,Meloni:2011fx}. The group $S_4$ admits two generators $S$ and $T$ satisfying
\be 
S^2=(ST)^3=T^4=1\,.
\ee
Its irreducible representations are two singlets $1$ and $1'$, a doublet 2 and two triplets $3$ and $3'$.
In one of the two triplet representations of $S_4$, $S$ and $T$ can be chosen as:
\be
T=\left(
\begin{array}{ccc}
-1&0&0\\
0&-i&0\\
0&0&i
\end{array}
\right)\,,\qquad\qquad
S=\left(
\begin{array}{ccc}
0&-\frac{1}{\sqrt{2}}&-\frac{1}{\sqrt{2}}\\
-\frac{1}{\sqrt{2}}&\frac{1}{2}&-\frac{1}{2}\\
-\frac{1}{\sqrt{2}}&-\frac{1}{2}&\frac{1}{2}
\end{array}
\right)\,.
\ee
In the class of models considered here the electroweak SU(2) lepton doublets $l$ transform as a triplet 3, the electroweak singlets
$e^c$, $\mu^c$ and $\tau^c$ transform as singlets and the electroweak Higgs doublets $H_{u,d}$ are invariant. If present, the right-handed neutrinos transform as a triplet 3.
Similarly to the previous class of models, at the LO and in the SUSY limit the VEVs of $\Phi_e$ and $\Phi_\nu$ are determined by two decoupled equations and satisfy
\be
T '\langle\Phi_e\rangle = \langle\Phi_e\rangle\,,\qquad\qquad
S\langle\Phi_\nu\rangle = \langle\Phi_\nu\rangle\qquad\qquad
\text{[\rm LO]}\,.
\label{sandtS4}
\ee
The transformation $T'=i T$ generates a $Z_4$ subgroup of the flavour group $G_f$ \cite{Altarelli:2009gn}.
The charged lepton mass matrix $m_e$ is given by  
\be
m_e= m_e^{(0)}+\delta m_e^{(1)}+\ldots
\ee
with the LO contribution $m_e^{(0)}$ of the same type as the one considered before in eq.~(\ref{mezero}).
In this case, $\eta$ represents a small parameter that breaks the flavour symmetry $S_4$. At the LO the charged lepton mass matrix $m^{(0)}_e$, only depending on $\langle \Phi_e\rangle$, is diagonal and invariant under the action of the transformation $T'$:
\be
{T'}^\dagger m^{(0) \dagger}_e m_e^{(0)} T'=m^{(0)\dagger}_e m_e^{(0)}\,.
\ee
We have $G_e=Z_4$, generated by $T'$. The hierarchical pattern $y_e\ll y_\mu\ll y_\tau$ is reproduced by operators of increasing dimensions contributing to $y_\tau$, $y_\mu$ and $y_e$, respectively.
Also in this case we are including in $m_e^{(0)}$ all terms arising from the LO $\langle\Phi_e\rangle$, independently from the dimensionality of the operators that contribute to the charged lepton mass matrix.

In the specific model of Ref.~\cite{Altarelli:2009gn} a See-Saw mechanism produces a mass matrix for the light neutrinos $m_\nu$, given by:
\be
m_\nu=m_\nu^{(0)}
+\delta m_\nu^{(1)}+\ldots
\ee
where
\be
m_\nu^{(0)}=\left(
\begin{array}{ccc}
x&y&y\nn\\
y&z&x-z\nn\\
y&x-z&z
\end{array}
\right)\,.
\label{mnulos4}
\ee
The parameters $x$, $y$ and $z$ are quadratic in $v_u$, the VEV of $H_u$ and inversely proportional to the scale associated with the violation
of the total lepton number. The LO term $m_\nu^{(0)}$ is invariant under the $Z_2\times Z_2$ symmetry generated by $S$, eq.~(\ref{sandtS4}), and by the transformation $A_{23}$ of eq.~(\ref{a23}):
\be
S^T m_\nu^{(0)} S=m_\nu^{(0)}\,,\qquad\qquad
A_{23}^T m_\nu^{(0)} A_{23}=m_\nu^{(0)}\,.
\ee
Also in this case the $Z_2$ symmetry represented by the matrix $A_{23}$ is an accidental symmetry. 
The matrix $m_\nu^{(0)}$ of eq.~(\ref{mnulos4}) is the most general one invariant under both $S$ and $A_{23}$. In the realization of Ref.  \cite{Altarelli:2009gn} the parameters $x$, $y$ and $z$ are related by
\be
z=x-\frac{y^2}{x}\,.
\ee 
At the LO $m_e^{(0)}$ is diagonal while $m_\nu^{(0)}$ is diagonalized by 
\be
U_{BM}=
\left(
\begin{array}{ccc}
\sqrt{\frac{1}{2}}&-\sqrt{\frac{1}{2}}&0\\
\frac{1}{2}&\frac{1}{2}&-\sqrt{\frac{1}{2}}\\
\frac{1}{2}&\frac{1}{2}&\sqrt{\frac{1}{2}}\\
\end{array}
\right)\,.
\ee
This holds for any value of the parameters $x$, $y$ and $z$ since $U_{BM}$ is the matrix that simultaneously diagonalizes $S$ and $A_{23}$.

The departure from the bimaximal mixing depends on the subleading contributions $\delta m_e^{(1)}$, $\delta m_\nu^{(1)}$.
In the model of Ref.~\cite{Altarelli:2009gn}, such contributions are not generic. At the NLO the corrections to both the neutrino and the charged lepton sector are controlled by $\langle\Phi_\nu\rangle$ 
which preserves the LO alignment. The NLO correction to the charged lepton mass matrix is no longer invariant under $T'$ and is of the following type:
\be
\delta m_e^{(1)}=
v_d
\left(
\begin{array}{ccc}
\cO(y_e)&\cO(y_e)&\cO(y_e)\nn\\
\cO(y_\mu)&\cO(y_\mu)&0\nn\\
\cO(y_\tau)&0&\cO(y_\tau)
\end{array}
\right)\eta~\xi\,,
\ee
where $\xi$ is a small adimensional parameter given by the ratio between a flavon of the $\Phi_\nu$ sector and $\Lambda$. The transformation needed to diagonalize $m_e$ is $V_e^T m_e U_e=m_e^{diag}$ where, to first order
in $\xi$
\beq
U_e=
\left(
\begin{array}{ccc}
1							&c^e_{12}\,\xi		&c^e_{13}\,\xi\\
-c^{e*}_{12}\,\xi			&1					&0\\
-c^{e*}_{13}\,\xi			&0					&1\\
\end{array}
\right)\,,
\eeq
where $c^e_{ij}$ are complex number with absolute value of order one. In the neutrino sector after the inclusion of the NLO corrections the mass matrix has still the form of eq.~(\ref{mnulos4}) and is diagonalized by $U_{BM}$.
The lepton mixing is $U_e^\dagger U_{BM}$ and to first order in $\xi$ we have
\beq
\begin{aligned}
\delta_{CP}					&=\pi+\arg\left(c^e_{12}-c^e_{13}\right)\\
\sin\theta_{13}		&=\frac{1}{\sqrt{2}}\,{|c^e_{12}-c^e_{13}|}\,\xi\\
\sin^2\theta_{12}	&=\frac{1}{2}-\frac{1}{\sqrt{2}}\,{\cal R}e(c^e_{12}+c^e_{13})\,\xi\\
\sin^2\theta_{23}	&=\frac{1}{2}\,.
\end{aligned}
\label{sinNLOBM}
\eeq

\begin{figure}[h!]
 \centering
 \subfigure[$c^e_{13}\neq0$]
   {\includegraphics[width=8.3cm]{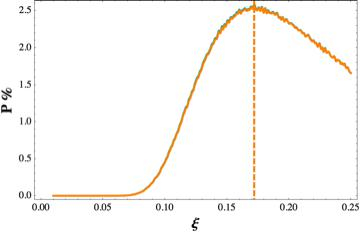}}
  \subfigure[$c^e_{13}=0$]
   {\includegraphics[width=7.7cm]{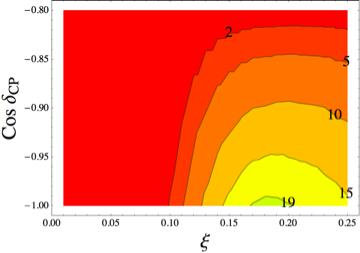}}
 \caption{{\bf {\boldmath$S_4$\unboldmath} Models.} In (a), the success rate as a function of the parameter $\xi$. The parameters $c^e_{12,13}$ have been taken as random complex numbers with absolute value following a Gaussian distribution around 1 with variance 0.5. In Cyan the NH and in Orange the IH. The value of $\xi$ that maximizes the success rate shown in the left plot is $\xi=0.172$, for both the NH and IH. In (b), the success rate for the case $c^e_{13}=0$ as a function of $\xi$ and $\cos \delta_{CP}$. The parameters $c^e_{12}$ is been taken as a random complex number with absolute value following a Gaussian distribution around 1 with variance 0.5. The success rate reaches its maximum, $\sim20\%$, for $\cos \delta_{CP}=-1$ and $\xi=0.18$. Only the NH case is shown. The IH is similar.}
 \label{fig:SuccessRatesBM}
\end{figure}

To properly correct the BM value of the solar angle to agree with the data, $\xi$ is expected to be $\cO(\lambda_C)$. Studying the success rate of having all three mixing angles inside the corresponding $3\sigma$ ranges, we find that it is maximized for both the NH and IH when $\xi=0.172$, as shown in Fig.~\ref{fig:SuccessRatesBM}(a). In this case, the maximal success rate of about $2.6 \%$ is particularly small. The problem for this model is not to reproduce $\theta_{13}$ but rather to calibrate the correction to $\theta_{12}$ for it to fall in its allowed window: we see from Fig.~\ref{fig:SuccessRatesBM}(a) that most of the scanning points spread out in a large interval of $\sin^2\theta_{12}$ between $\sim 0.2$ and $\sim 0.8$.
It may be interesting in this case to explore the possibility that one of the charged lepton mixing angles is dominant. For this to occur naturally an additional dynamical input would be needed. For the specific case $c^e_{13}=0$, we get a even more predictive correlation (still dependent, through $\delta_{CP}$, on the $c^e_{12}$ phase) among the solar and the reactor angle:
\beq
\sin^2\theta_{12}=\frac{1}{2}+\sin\theta_{13}\,\cos\delta_{CP}+\cO(\sin^2\theta_{13})\,.
\label{BM_PreciseCorrelation}
\eeq
In this case, we study the dependence of the success rate from both $\xi$ and the Dirac CP phase, in terms of $\cos\delta_{CP}$: being dependent on two parameters, this success rate cannot be directly compared with the previous one, function of only $\xi$; a trustworthy comparison requires to average among all the values of the success rate for a fixed $\xi$. The corresponding plot for the NH case is shown in Fig.~\ref{fig:SuccessRatesBM}(b): the success rate reaches its maximum, $\sim20\%$, for $\xi=0.18$ and $\cos \delta_{CP}=-1$, for both the NH  and IH cases.  Only the NH case is shown in Fig.~\ref{fig:SuccessRatesBM}(b), while the IH is similar. The alternative possibility that $c^e_{12}=0$ would lead to similar results.

\begin{figure}[h!]
 \centering
 \includegraphics[width=9cm]{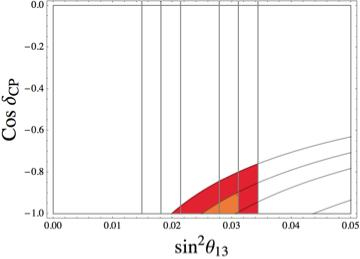}
 \caption{ {\bf {\boldmath$S_4$\unboldmath} Models.} Contour plot for $1,2,3\sigma$ values of $\sin^2\theta_{12}$ in the parameter space $\sin^2\theta_{13}$--$\cos\delta_{CP}$ according to the values for the NH in Table~\ref{tab:data}. The Orange and Red regions refers to the data at $2\sigma$ and $3\sigma$, respectively. No region corresponding to the data at $1\sigma$ is present.}
 \label{fig:CorrelationsBM}
\end{figure}

It is interesting to investigate the origin of the preference for $\delta_{CP}=-1$ shown in Fig.~\ref{fig:SuccessRatesBM}(b). In Fig.~\ref{fig:CorrelationsBM}, we graphically present the correlation in Eq.~(\ref{BM_PreciseCorrelation}), neglecting the $\cO(\sin^2\theta_{13})$ terms. The predicted value of $\sin^2\theta_{12}$ agrees with the experimental one
only when $\cos\delta_{CP}$ is very close to $-1$. Although this correlation is modified once subleading contributions from both the neutrino and the charged lepton sectors are taken into account, still it will provide a strong test for such models.

\begin{figure}[h!]
 \centering
 \subfigure[$c_{13}\neq0$]
   {\includegraphics[width=7.7cm]{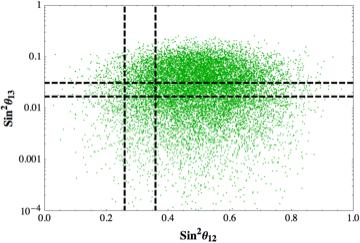}}
 \subfigure[$c_{13}=0$]
   {\includegraphics[width=7.7cm]{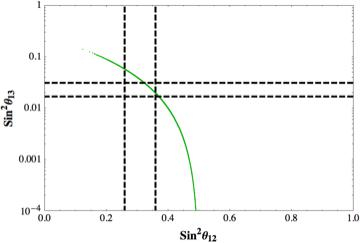}}
 \caption{{\bf {\boldmath$S_4$\unboldmath} Models.} $\sin^2\theta_{13}$ as a function of $\sin^2\theta_{12}$ is plotted, following eq.~(\ref{sinNLOBM}). The dashed-black lines represent the $3\sigma$ values for the mixing angles from the Fogli {\it et al.} fit \cite{Fogli:2012ua}. Only the NH data sets is shown. On the left, the parameters $c^e_{ij}$ are random complex numbers with absolute values following a Gaussian distribution around 1 with variance 0.5, while $\xi=0.172$. In the right, $c^e_{13}=0$, $\xi=0.18$ and $\cos \delta_{CP}=-1$.}
 \label{fig:Sin12qvsSin13qBM_NIH}
\end{figure}

We analyze quantitatively the expressions in eq.~(\ref{sinNLOBM}) and their correlation in Fig.~\ref{fig:Sin12qvsSin13qBM_NIH}(a), where $c^e_{12,13}$ have been taken as random complex numbers with absolute value following a Gaussian distribution around 1 with variance 0.5, while $\xi=0.172$. In Fig.~\ref{fig:Sin12qvsSin13qBM_NIH}(b), we analyze the specific case $c^e_{13}=0$, where $\xi=0.18$ and $\cos \delta_{CP}=-1$. In Fig.~\ref{fig:Sin12qvsSin13qBM_NIH}, only the NH case is shown. The IH case is similar. Comparing Fig.~\ref{fig:SuccessRatesBM}(a) with Figs.~\ref{fig:SuccessRatesTB} and \ref{fig:SuccessRatesTBLin}, we can see that these $S_4$ models are strongly disfavoured with respect to the $A_4$ ones, and especially with respect to the special $A_4$ models.

In conclusion, from the point of view of reproducing the observed values of the mixing angles in a natural way, the $A_4$ models of the Lin type provide a most efficient solution. In the next sections we will study the performance of the different models for LFV processes.

%%%%%%%%%%%%%%%%%%%%%%%%%%%%%%%%%%%%%%%%%%%%%%%%%%%%%%%%%%%%%%%%%
%%%%%%%%%%%%%%%%%%%%%%%%%   3.  Lepton Flavour Violation      
%%%%%%%%%%%%%%%%%%%%%%%%%%%%%%%%%%%%%%%%%%%%%%%%%%%%%%%%%%%%%%%%%
\section{Lepton Flavour Violation}
\label{sec:LFVformalism}

In this section we discuss the implications on LFV processes of the three classes of models described in the previous section. In particular, given the stringent upper bound on $BR(\mu\to e \gamma)$, we focus on radiative lepton decays. By working in the super-CKM basis, where all kinetic terms are canonical and lepton mass matrices have been diagonalized through unitary transformations acting on the whole supermultiplets,
the only sources of LFV are the off-diagonal terms of the slepton mass matrices. In our models these terms are much smaller than the corresponding diagonal entries and we can use the mass-insertion (MI) approximation \cite{Hall:1985dx,Masina:2002mv,Paradisi:2005fk} to illustrate the qualitative behavior of the model predictions. The quantitative results shown in our plots have been obtained by using complete one-loop results, which can be found for example in Refs.~\cite{Hisano:1995nq,Hisano:1995cp,Hisano:2001qz,Fukuyama:2005bh,Arganda:2005ji,Isidori:2007jw,Endo:2008um}. The normalized branching ratios $R_{ij}$ for the LFV transitions $l_i\to l_j \gamma$
\be
R_{ij}=\frac{BR(l_i\to l_j\gamma)}{BR(l_i\to l_j\nu_i{\bar \nu_j})}
\ee
can be written as
\be
R_{ij}= \frac{48\pi^3 \alpha}{G_F^2 m_{SUSY}^4}
\left(\vert A_L^{ij} \vert^2+\vert A_R^{ij} \vert^2 \right)\,.
\label{rij}
\ee
At the LO in the MI approximation, the amplitudes $A_L^{ij}$ and $A_R^{ij}$ are given by:
\be
\begin{aligned}
A_L^{ij}&=a_{LL} (\delta_{ij})_{LL} + a_{RL} \frac{m_{SUSY}}{m_i} (\delta_{ij})_{RL}\\
A_R^{ij}&=a_{RR} (\delta_{ij})_{RR} + a_{LR} \frac{m_{SUSY}}{m_i} (\delta_{ij})_{LR}
\label{ALAR}
\end{aligned}
\ee
where $m_i$ are the charged fermion masses and $a_{CC'}$ $(C,C'=L,R)$ are dimensionless functions of the SUSY parameters $m_{SUSY}$, $M_{1,2}$, $\mu$, $\tan\beta$,
renormalized at the electroweak scale. Here a common value in the diagonal entries of both LL and  RR blocks of the slepton mass matrices has been assumed at the electroweak scale and $m_{SUSY}$ denotes the average mass.
Such assumption, often made at the cut-off scale in constrained versions of the MSSM, is spoiled by running effects and does not hold any more at the electroweak scale. We will 
discuss this effect in the next subsection. The simplified framework considered here is sufficient to correctly describe the relation between $R_{ij}$ and the expansion parameters $\xi$ and $\xi'$.
In our conventions the explicit expression of $a_{CC'}$ is given in appendix A. Their typical size is one tenth of $g^2/(16\pi^2)$, $g$ being the SU(2)$_L$ gauge coupling constant. 
To appreciate the relative weights of the contributions in eq.~(\ref{ALAR}), we list in Tab.~\ref{table_aCC} the expressions and the numerical values of the functions $a_{CC'}$, in the limit $\mu=M_{1,2}=m_{SUSY}$. 
As one can see, in this limit the dominant coefficient is  ${a}_{LL}$, which is larger than $a_{RL}=a_{LR}$ by a factor $7\div 54$, 
and larger than $a_{RR}$ by a factor $4\div 14$, depending on  $\tan\beta=2\div 15$. This range of $\tan\beta$ is taken as representative for the models under consideration.
More precisely, the parameter $\tan\beta$ is related to the expansion parameter $\eta$, the mass of the $\tau$ lepton and the $\tau$ Yukawa coupling $y_\tau$, by:
\beq
\tan\beta\approx \dfrac{|y_\tau|\,\eta\,v_{EW}}{\sqrt2\,m_\tau}\,,
\label{tanbeta}
\eeq 
where $v_{EW}\approx246\GeV$ is the EW Higgs VEV. For typical $A_4$ models the parameter $\eta\approx\xi$ is of order 0.1. For special $A4$ model, $\eta\approx\xi$ is smaller than $\xi'$ and we will use the range
$0.007\lesssim\eta\lesssim0.05$. For the $S_4$ models, we have $\eta\approx 0.08$ to correctly fit the charged lepton masses. Requiring $y_\tau$ to be of order one, $1/3 \lesssim |y_\tau| \lesssim 3$, 
we have the following allowed ranges for $\tan\beta$:
\beq
\begin{aligned}
&3 \lesssim \tan\beta \lesssim 30\qquad&& \text{Typical}~ A_4\\
&2 \lesssim \tan\beta \lesssim 15\qquad&& \text{Special}~ A_4\\
&3 \lesssim \tan\beta \lesssim 24\qquad&&  S_4
\end{aligned}
\eeq

\begin{table}[!ht]
\centering
\begin{math}
\begin{array}{|c|l|c|}
\hline
&&  \\[-3mm]
a_{LL}&\dd\frac{1}{240}\frac{g^2}{16 \pi^2}\left[1-3\left(1+4 c\right)\tan^2\theta_W+4 \Big(4+5\tan^2\theta_W\Big)\tan\beta\right]&+(2.0\div16.3) \\[4mm]
\hline
&&   \\[-3mm]
a_{RL}=a_{LR}&\dd\frac{1}{12}\frac{g^2}{16 \pi^2}\tan^2\theta_W&0.30 \\[4mm]
\hline 
&&      \\[-3mm]
a_{RR}&\dd\frac{1}{60}\frac{g^2}{16 \pi^2}\tan^2\theta_W\left[-3-3 c-\tan\beta\right]&-(0.5\div1.2) \\[4mm]
\hline
\end{array}
\end{math}
\caption{\label{table_aCC} Coefficients $a_{CC'}$ characterizing the transition amplitudes for $\mu\to e \gamma$, $\tau\to e \gamma$ and $\tau\to \mu\gamma$, in the MI approximation and by taking the limit the $\mu=M_{1,2}=m_{SUSY}$.  Numerical values are given in units of $g^2/(192 \pi^2)$ and using $\sin^2\theta_W=0.23$, $c=1$, and $\tan\beta=2\div 15$. The parameter $c$ is a model dependent quantity of order one.}
\end{table}

Finally, $(\delta_{ij})_{CC'}$ parametrize the MIs and are defined as:
\be
(\delta_{ij})_{CC'}=\frac{(\hat{m}^2_{eCC'})_{ij}}{m^2_{SUSY}},
\label{MI}
\ee
where the different blocks of the slepton mass matrices $\hat{m}^2_{eCC'}$ are evaluated in the super-CKM basis, denoted by the hat. There are two main types of contributions to the MIs $(\delta_{ij})_{CC'}$: the first one comes from local operators, 
bilinear in the slepton fields, with insertions of the flavons $\Phi_{e,\nu}$. They are invariant under the flavour symmetry and represent the counterpart, in the sparticle sector, of the operators 
that generate lepton masses in the theory. After the breaking of the flavour symmetry, we get slepton mass matrices expanded up to a certain order in $\langle\Phi_{e,\nu}\rangle/\Lambda$,
depending on the highest dimensionality of the operators included:
\be
\hat{m}^2=\hat{m}^2_0+\delta\hat{m}^2_1+\delta\hat{m}^2_2+...
\ee
with $\delta\hat{m}^2_p$ of order $(\langle\Phi_{e,\nu}\rangle/\Lambda)^p$.

The second contribution to the MIs in eq. (\ref{MI}) comes from the renormalization group evolution (RGE). 
When neutrino masses are generated by the See-Saw mechanism, the evolution spans two main regions. The first one goes from the cut-off scale $\Lambda$
where slepton masses are generated\footnote{Depending on the specific type of SUSY breaking mechanism slepton mass generation can occur at a scale smaller than $\Lambda$. Here we assume that slepton masses are
produced at the highest scale, like in gravity mediated SUSY breaking scenarios.}, down to the lightest right-handed neutrino mass $M_1$. In our models the mass scale associated to right-handed neutrinos 
coincides with the flavour symmetry breaking scale $\langle\Phi_{\nu}\rangle$ and, in a leading logarithmic approximation, we get a contribution to the MIs
proportional to $y^2/(16\pi^2)\times \log(\langle\Phi_{\nu}\rangle/\Lambda)$, $y$ representing a typical neutrino Yukawa coupling. 
Below $M_1$ the right-handed neutrinos decouple and the running is only affected by the lightest degrees of freedom, those of
the MSSM. This part of the RGE occurs whether or not neutrino masses originate from the See-Saw mechanism and gives rise to contributions to the MIs
proportional to $y_\tau^2/(16\pi^2)\times \log(\langle\Phi_{\nu}\rangle/m_{SUSY})$, $y_\tau$ denoting the $\tau$ Yukawa coupling.

We now discuss in turn the different contributions to the MIs for
the three classes of models.

%%%%%%%%%%%%%%%%%%%%%%%%%   3.1 MI from Local Operators      
\subsection{Mass Insertions from Local Operators}

Local operators giving rise to slepton masses can be constructed with standard techniques, see Refs.~\cite{Hamaguchi:2002vi,Mondragon:2007af,Kifune:2007fj,Ishimori:2008ns,Feruglio:2008ht,Ishimori:2008au,Ishimori:2009ew,Feruglio:2009iu,Feruglio:2009hu,Hagedorn:2009df,Merlo:2011hw,Chakrabortty:2012vp}. The inclusion of non-renormalizable operators with insertions of the flavon fields
$\Phi_{e,\nu}$ also affects kinetic terms both in the fermion and sfermion sector, providing an additional source of flavour violation. 
Here we list the slepton mass matrices in the super-CKM basis where kinetic terms have been set in the canonical form by means of appropriate transformations
and where fermion mass matrices have been made diagonal through unitary transformations acting on the whole supermultiplet. 
These slepton masses refer to specific models, taken as representative of the classes analyzed above. For typical $A_4$ models we will refer to the construction in Ref.  \cite{Altarelli:2005yx},
for special $A_4$ models we will consider the model of Ref. \cite{Lin:2009bw} and finally $S_4$ models are exemplified by the model in Ref. \cite{Altarelli:2009gn}.
The results given below have been obtained under the assumption that the underlying parameters are real.

%%%%%%%%%%%%%%%%%%%%%%%%%   3.1.1  Typical $A_4$ models  
\boldmath
\subsubsection{Typical $A_4$ Models}
\label{sec:LFVTypicalA4}
\unboldmath
For typical $A_4$ models we get
\be
\hat{m}_{LL}^2=\left(\begin{array}{ccc}
1+\cO(\xi)		& \cO(\xi^2)			& \cO(\xi^2)\\[0.1in]
\cO(\xi^2)		&1+\cO(\xi)			& \cO(\xi^2)\\[0.1in]
\cO(\xi^2)		& \cO(\xi^2)			& 1+\cO(\xi)
\end{array}\right)m_{SUSY}^2+\ldots
\label{TA4slm1}
\ee
\be
\hat{m}_{RR}^2= \left( \begin{array}{ccc}
                 \cO(1)
                                & \dd\frac{m_e}{m_\mu}\cO(\xi)
                                & \dd\frac{m_e}{m_\tau}\cO(\xi)\\[0.1in]
                 \dd\frac{m_e}{m_\mu}\cO(\xi)
                                & \cO(1)
                                &  \dd\frac{m_\mu}{m_\tau} \cO(\xi)\\[0.1in]
                \dd\frac{m_e}{m_\tau} \cO(\xi)
                                & \dd \frac{m_\mu}{m_\tau} \cO(\xi)
                                & \cO(1)
		\end{array}\right) \, m_{SUSY}^2+\ldots
		\label{TA4slm2}
\ee             
\be
\hat{m}_{RL}^2= \left( \begin{array}{ccc}     
				\cO(m_e)
                                &  m_e \cO(\xi)
                                &  m_e \cO(\xi)\\[0.1in]
                m_\mu \cO(\xi^2)
                                &  \cO(m_\mu)
                                & m_\mu \cO(\xi)\\[0.1in]
                 m_\tau \cO(\xi^2)
                                & m_\tau \cO(\xi^2)
                                & \cO(m_\tau)
\end{array}\right) \, m_{SUSY}+\ldots
\label{TA4slm3}
\ee                 
where dots stand for negligible SUSY contributions. In $\hat{m}_{RL}^2$ we have neglected a contribution which arises if the $F$ component of the flavon supermultiplets
acquires a VEV. Such a contribution depends on the SUSY breaking mechanism and vanishes under mild assumptions \cite{Ross:2002mr,Antusch:2008jf,Feruglio:2009iu}. A similar contribution will be neglected also in the special $A_4$ and in the $S_4$ models
discussed below.
We have
\beq
\begin{aligned}
(\delta_{ij})_{LL}&=\cO(\xi^2)\,,\qquad						&	(\delta_{ij})_{RL}&=\frac{m_i}{m_{SUSY}}\cO(\xi^2)\,,\\
(\delta_{ij})_{RR}&=\frac{m_j}{m_i}\cO(\xi)\,,\qquad	&  (\delta_{ij})_{LR}&=\frac{m_j}{m_{SUSY}}\cO(\xi)\,,
\end{aligned}
\eeq       
and 
\beq
\begin{aligned}
A^{ij}_L&=a_{LL} \cO(\xi^2)+a_{RL}\cO(\xi^2)\,,\\
A^{ij}_R&=a_{RR} \frac{m_j}{m_i} \cO(\xi)+a_{RL}\frac{m_j}{m_i} \cO(\xi)\,.
\end{aligned}
\eeq
From $\xi\approx 0.1$ and the numerical values of $a_{CC'}$ we see that the amplitude $A^{ij}_L$ is the dominant one. We approximately have
\beq
R_{ij}\simeq \frac{48\pi^3 \alpha}{G_F^2 m_{SUSY}^4}
\left| a_{LL} +a_{RL} \right|^2\,\cO(\xi^4)\,,
\eeq
and we expect the branching ratios of the three transitions to be of the same order of magnitude:  
\beq
R_{\mu e}\approx R_{\tau\mu} \approx R_{\tau e}\,,
\eeq
at variance with the predictions of most of the other models, where, for instance, $R_{\mu e} /R_{\tau \mu}$ can be much smaller than one \cite{Cirigliano:2005ck,Davidson:2006bd,Grinstein:2006cg,Alonso:2011jd}.

%%%%%%%%%%%%%%%%%%%%%%%%%   3.1.2  Special $A_4$ models  
\boldmath
\subsubsection{Special $A_4$ Models}
\label{sec:LFVSpecialA4}
\unboldmath

For the special $A_4$ model in Ref.  \cite{Lin:2009bw} we get :
\begin{equation}
\hat{m}_{LL}^2=
\left( \begin{array}{ccc}
                1+ \cO(\xi'^2)  &  \cO(\xi'^2) & \cO(\xi'^2) \\[0.1in]
                \cO(\xi'^2) 	& 1+ \cO(\xi'^2) & \cO(\xi'^2)\\[0.1in]
                \cO(\xi'^2) 	& \cO(\xi'^2) & 1+\cO(\xi'^2) 
\end{array}\right) \, m_{SUSY}^2+\ldots
\label{SA4slm1}
\end{equation}
\be
\hat{m}_{RR}^2= \left( \begin{array}{ccc}
                 \cO(1)
                                & \dd\frac{m_e}{m_\mu}\cO(\xi'^2)
                                & \cO(\xi'^3)\\[0.1in]
                 \dd\frac{m_e}{m_\mu}\cO(\xi'^2)
                                & \cO(1)
                                &  \dd\frac{m_\mu}{m_\tau} \cO(\xi'^2)\\[0.1in]
                \cO(\xi'^3)
                                & \dd \frac{m_\mu}{m_\tau} \cO(\xi'^2)
                                & \cO(1)
		\end{array}\right) \, m_{SUSY}^2+\ldots
		\label{SA4slm2}
\ee        
\be
\hat{m}_{RL}^2= \left( \begin{array}{ccc}     
\cO(m_e)
                                &  m_e \cO(\xi'^2)
                                &  m_e \cO(\xi'^2)\\[0.1in]
                m_\mu \cO(\xi'^2)
                                &  \cO(m_\mu)
                                & m_\mu \cO(\xi'^2)\\[0.1in]
                 m_\tau \cO(\xi'^2)
                                & m_\tau \cO(\xi'^2)
                                & \cO(m_\tau)
\end{array}\right) \, m_{SUSY}+\ldots
\label{SA4slm3}
\ee                 

We have
\beq
\begin{aligned}
(\delta_{ij})_{LL}&=\cO(\xi'^2)\,,\\
(\delta_{21})_{RR}&=\frac{m_e}{m_\mu}\cO(\xi'^2)&(\delta_{32})_{RR}&=\frac{m_\mu}{m_\tau}\cO(\xi'^2)~~~(\delta_{31})_{RR}=\cO(\xi'^3)\,,\\
(\delta_{ij})_{RL}&=\frac{m_i}{m_{SUSY}}\cO(\xi'^2)&(\delta_{ij})_{LR}&=\frac{m_j}{m_{SUSY}}\cO(\xi'^2)\,
\end{aligned}
\eeq       
and 
\beq
\begin{aligned}
A^{ij}_L&=a_{LL} \cO(\xi'^2)+a_{RL}\cO(\xi'^2)\,,\\
A^{ij}_R&=
\begin{cases}
a_{RR} \dd\frac{m_j}{m_i} \cO(\xi'^2)+a_{RL}\dd\frac{m_j}{m_i} \cO(\xi'^2)\,\qquad 	&(ij=21,32)\\
a_{RR} \cO(\xi'^3)+a_{RL}\dd\frac{m_j}{m_i} \cO(\xi'^2)\,\qquad								&(ij=31)\,.
\end{cases}
\end{aligned}
\eeq
Neglecting the subdominant contribution from the $A^{ij}_R$ amplitude, we have
\beq
R_{ij}\simeq \frac{48\pi^3 \alpha}{G_F^2 m_{SUSY}^4}
\left| a_{LL} +a_{RL} \right|^2\,\cO(\xi'^4)\,,
\eeq
and we expect the branching ratios of the three transitions to be of the same order of magnitude:
\beq
R_{\mu e}\approx R_{\tau\mu} \approx R_{\tau e}\,.
\eeq

%%%%%%%%%%%%%%%%%%%%%%%%%   3.1.3  $S_4$ Models  
\boldmath
\subsubsection{$S_4$ Models}
\label{sec:LFVS4}
\unboldmath

For $S_4$ models we get:
\be
\hat{m}_{LL}^2=\left(\begin{array}{ccc}
1+\cO(\xi)& \cO(\xi)& \cO(\xi)\\[0.1in]
\cO(\xi)&1+\cO(\xi)& \cO(\xi^2)\\[0.1in]
\cO(\xi)& \cO(\xi^2)& 1+\cO(\xi)
\end{array}\right)m_{SUSY}^2+\ldots
\label{S4slm1}
\ee
\be
\hat{m}_{RR}^2= \left( \begin{array}{ccc}
                 \cO(1)
                                & \dd\frac{m_e}{m_\mu}\cO(\xi)
                                & \dd\frac{m_e}{m_\tau}\cO(\xi)\\[0.1in]
                 \dd\frac{m_e}{m_\mu}\cO(\xi)
                                & \cO(1)
                                &  \dd\frac{m_\mu}{m_\tau} \cO(\xi^2)\\[0.1in]
                \dd\frac{m_e}{m_\tau} \cO(\xi)
                                & \dd \frac{m_\mu}{m_\tau} \cO(\xi^2)
                                & \cO(1)
		\end{array}\right) \, m_{SUSY}^2+\ldots
		\label{S4slm2}
\ee             
\be
\hat{m}_{RL}^2= \left( \begin{array}{ccc}     
\cO(m_e)
                                &  m_e \cO(\xi)
                                &  m_e \cO(\xi)\\[0.1in]
                m_\mu \cO(\xi)
                                &  \cO(m_\mu)
                                & m_\mu \cO(\xi^2)\\[0.1in]
                 m_\tau \cO(\xi)
                                & m_\tau \cO(\xi^2)
                                & \cO(m_\tau)
\end{array}\right) \, m_{SUSY}+\ldots
\label{S4slm3}
\ee        
We have
\beq
\begin{aligned}
(\delta_{ij})_{LL}&=\cO(\xi^p)\,,\qquad						&	(\delta_{ij})_{RL}&=\frac{m_i}{m_{SUSY}}\cO(\xi^p)\,,\\
(\delta_{ij})_{RR}&=\frac{m_j}{m_i}\cO(\xi^p)\,,\qquad	&  (\delta_{ij})_{LR}&=\frac{m_j}{m_{SUSY}}\cO(\xi^p)\,
\end{aligned}
\eeq       
where $p=1$ when $ij=21,31$ and $p=2$ when $ij=32$
\beq
\begin{aligned}
A^{ij}_L&=a_{LL} \cO(\xi^p)+a_{RL}\cO(\xi^p)\,,\\
A^{ij}_R&=a_{RR} \frac{m_j}{m_i} \cO(\xi^p)+a_{RL}\frac{m_j}{m_i} \cO(\xi^p)\,.
\end{aligned}
\eeq
Neglecting the subdominant contribution from the $A^{ij}_R$ amplitude, we have
\beq
R_{ij}\simeq \frac{48\pi^3 \alpha}{G_F^2 m_{SUSY}^4}
\left| a_{LL} +a_{RL} \right|^2\times
\begin{cases}
\cO(\xi^2)\qquad		&(ij=21,31)\\
\cO(\xi^4)\qquad	&(ij=32)\,,
\end{cases}
\eeq
with a suppression of the rate $\tau\to \mu\gamma$ relative to $\mu\to e \gamma$ and $\tau\to e \gamma$ by a factor of $\xi^2$:
\beq
R_{\tau\mu}\ll R_{\mu e} \approx R_{\tau e}\,.
\eeq
Before a more quantitative illustration of these results, we discuss the effects induced by the 
RGE from the scale of flavour symmetry breaking down to the electroweak scale.

%%%%%%%%%%%%%%%%%%%%%%%%%   3.2 MI from Low-Energy RGE
\subsection{Mass Insertions from low-energy RGE}

This set of corrections is common to all models irrespective of the assumed existence of RH neutrinos, i.e.~both with or without See-Saw.
The running of the slepton mass parameters from the scale $\langle\Phi_{e,\nu}\rangle$ down to the scale $m_{SUSY}$ can be estimated in a leading logarithmic approximation.
The largest effect is a correction to the matrices $\hat{m}^2_{LL}$ and $\hat{m}^2_{RR}$ coming from electroweak gauge interactions and is proportional to the identity matrix in flavour space. 
One finds that the diagonal elements increase in the running from the cutoff scale down to the electroweak scale.
This effect is taken into account in our numerical study. As for the off diagonal entries the largest corrections are proportional to the square of the $\tau$ Yukawa coupling.
Since $y_\tau^2/(16\pi^2)\approx 3\times(10^{-6}\div 10^{-4})$ for $\tan\beta=(2\div 15)$, even in the presence of the large factor $\log(\langle\Phi_{e,\nu}\rangle/m_{SUSY})\approx 30$,
these corrections are negligibly small compared to the contribution from the local operators discussed above. We can conclude that the corrections to the off-diagonal entries of the soft mass matrices induced by the RGE from $\langle\Phi_{e,\nu}\rangle$ down to $m_{SUSY}$ are either negligible or could be absorbed in the parametrization given in the previous section.

The present bound on the branching ratio of $\mu\to e \gamma$ from the MEG collaboration \cite{Adam:2011ch}, $BR(\mu\to e\gamma)<2.4\times10^{-12}$,
leads to strong constraints on the parameter space of the models that we are considering in this work.
The supersymmetric parameters that are not constrained by the flavour symmetry, such as the soft SUSY mass scales and the gaugino and Higgs(ino) sectors, are fixed by our choice of a SUGRA framework: $m_0$ and $M_{1/2}$ are the common masses of scalar particles and gauginos at the GUT scale. Thus, at the scale $\Lambda = 2 \times 10^{16}\GeV$,
\beq
M_1(\Lambda) = M_2(\Lambda) = M_{1/2}\,,
\eeq
where $M_i$ are the $SU(2)\times U(1)$ gaugino masses. The effects of the RG running lead at low energies to the following masses for the gauginos
\beq
M_1(m_W)\simeq\dfrac{\alpha_1(m_W)}{\alpha_1(\Lambda)}M_1(\Lambda)\qquad\qquad
M_2(m_W)\simeq\dfrac{\alpha_2(m_W)}{\alpha_2(\Lambda)}M_2(\Lambda)\,,
\eeq
where $\alpha_i=g_i^2/4\pi$ ($i=1,2$) and $\alpha_1(\Lambda)=\alpha_2(\Lambda)\simeq 1/25$.
We have seen that, among the RG running effects on the soft mass terms, only those from the electroweak gauge interactions are relevant.
At the cut-off scale the $LL$ and $RR$ blocks of the slepton mass matrices are given by the eqs. (\ref{TA4slm1})-(\ref{TA4slm3}), (\ref{SA4slm1})-(\ref{SA4slm3}) and (\ref{S4slm1})-(\ref{S4slm3}) with $m_{SUSY}$ identified with $m_0$. 
The RGE due to electroweak gauge interactions leaves the off-diagonal entries essentially unaffected, while the diagonal elements at the weak scale,
denoted by $m_{L,R}^2(m_W)$, are given by:
\beq
\begin{aligned}
m_L^2(m_W)\simeq&\, m_0^2+0.54\, M_{1/2}^2 \,,\\
m_R^2(m_W)\simeq&\, m_0^2+0.15\, M_{1/2}^2 \,.
\end{aligned}
\eeq
Notice that this effect modifies the previous estimates of the $LL(RR)$ MIs by a factor $m_0^2/(m_0^2+0.54(0.15) M_{1/2}^2)$, thus providing an additional suppression when $M_{1/2}$ is larger than $m_0$.

Furthermore, the parameter $\mu$ is fixed through the requirement of correct electroweak symmetry breaking\footnote{The general definition of the parameter $\mu$ is
\beq
|\mu|^2=\dfrac{m_{H_d}^2-m_{H_u}^2\tan^2\beta}{\tan^2\beta-1}-\dfrac{1}{2}m_Z^2\,,
\eeq
that reduces to the expression in Eq.~(\ref{defmuSUGRA}) once considering that in the SUGRA framework the soft Higgs mass parameters are also given by $m_0$ at the high energy scale, $m_{H_u}^2 (\Lambda)=m_{H_d}^2 (\Lambda)=m_0^2$.}
\beq
|\mu|^2\simeq-\dfrac{m_Z^2}{2}+m_0^2\dfrac{1+0.5\tan^2\beta}{\tan^2\beta-1}+M_{1/2}^2\dfrac{0.5+3.5 \tan^2\beta}{\tan^2\beta-1}\;,
\label{defmuSUGRA}
\eeq
so that $\mu$ is determined by $m_0$, $M_{1/2}$  and $\tan\beta$ up to its sign. 
We recall that in our model the low energy parameter $\tan\beta$ is not a free parameter, as shown in Eq.~(\ref{tanbeta}). We compare the models at $\tan\beta=2$ and $\tan\beta=15$.  In our numerical analysis, we have assumed that the parameters on the diagonal of the slepton mass matrices $(m_{(e,\nu)LL}^2)_K$ and $(m_{eRR}^2)_K$ are positive in order to get positive definite square-masses and to avoid electric-charge breaking minima and further sources of electroweak symmetry breaking. The absolute value of the ${\cal O}(1)$ parameters is varied between $1/2$ and $2$. Furthermore, we have imposed the conditions that the lightest chargino has a mass larger than $100$ GeV and that the lightest neutralino is the lightest supersymmetric a particle (LSP).

\begin{figure}[h!]
 \centering
  \subfigure[$\tan\beta=2$ and $m_0=200\GeV$]
   {\includegraphics[width=7.5cm]{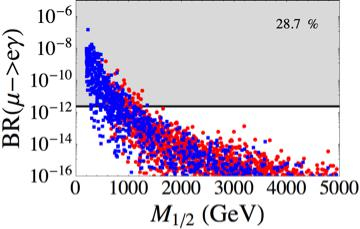}}
  \subfigure[$\tan\beta=2$ and $m_0=5000\GeV$]
   {\includegraphics[width=7.5cm]{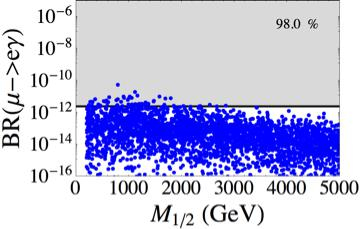}}
  \subfigure[$\tan\beta=15$ and $m_0=200\GeV$]
   {\includegraphics[width=7.5cm]{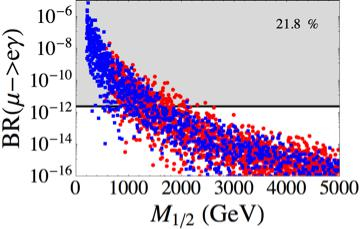}}
  \subfigure[$\tan\beta=15$ and $m_0=5000\GeV$]
   {\includegraphics[width=7.5cm]{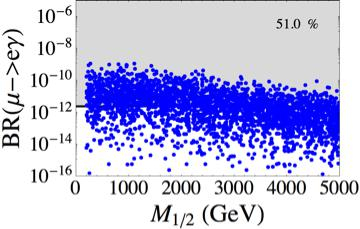}}
 \caption{{\bf Typical {\boldmath$A_4$\unboldmath} Models.} Scatter plots of $BR(\mu\to e \gamma)$ as a function of $M_{1/2}$, for different values of $\tan\beta$, and $m_0$. The parameter $\xi$ is chosen as $0.076$ in order to maximize the success rate of this model. The horizontal line shows the current MEG bound. For Blue (Red) points the LSP is the lightest neutralino (stau). The percentage in each plot refers to the number of Blue points that satisfy the MEG bound over the total number of points.}
 \label{fig:MEGTypicalA4Wein}
\end{figure}

The results for the typical $A_4$ models are illustrated in Fig.~\ref{fig:MEGTypicalA4Wein},
where the parameter $\xi$ is taken equal  to $0.076$, consistently with the analysis in Sec.~\ref{sec:TypicalA4}.
For $\tan\beta=2$ and $m_0=200\GeV$, Fig.~\ref{fig:MEGTypicalA4Wein}(a), almost all the points are excluded for $M_{1/2}\lesssim400\GeV$. The corresponding supersymmetric spectrum is rather light: for $M_{1/2}=400\GeV$, the lightest neutralino has a mass of $\sim156\GeV$, the lightest chargino of $\sim306\GeV$, the lightest LH charged slepton is in the range $[230,\,500]\GeV$ and the lightest RH charged slepton in the range $[160,\,350]\GeV$. These mass values are not yet excluded by the LHC (the limits in the EW sector are not very strong) and they only mark the lower edge of the region allowed by the $\mu \rightarrow e \gamma$ bounds. Increasing $m_0$ up to $5000\GeV$, Fig.~\ref{fig:MEGTypicalA4Wein}(b), we see that the MEG bound is well satisfied in this model for the whole plotted range of $M_{1/2}$. The corresponding supersymmetric spectrum is heavier: again for $M_{1/2}=400\GeV$, while the lightest neutralino and chargino have masses very similar to the previous case, $\sim158\GeV$ and $\sim315\GeV$, respectively, the lightest LH and RH charged slepton masses are much higher, being in the range $[3850,\,7070]\GeV$ and $[1800,\,6260]\GeV$, respectively. Increasing $m_0$ from $200$ GeV up to $5000$ GeV does not correspond to a uniform decrease of the branching ratio,
as a function of $M_{1/2}$. The plot with $m_0=5000$ GeV is much flatter compared to the plot where $m_0=200$ GeV. This is due to the approximate factor
$m_0^2/(m_0^2+0.54 M_{1/2}^2)$ entering the dominant $LL$ mass insertion, which, for the values of $M_{1/2}$ used in our plots, is sharply decreasing for $m_0=200$ GeV while it is slowly varying and close to one for $m_0=5000$ GeV. This is a general feature reproduced by all the models considered here.

Increasing the value of $\tan\beta$, Figs.~\ref{fig:MEGTypicalA4Wein}(c)-(d), the number of points below the MEG bound decreases, but the previous considerations are approximatively still valid: in particular notice that for $m_0=5000\GeV$, there are always points satisfying the MEG bound, even if the largest number of them falls in the excluded region, especially for smaller $M_{1/2}$. Since the dominant contribution to the branching ratio comes from the $LL$ mass insertion which is proportional to $\tan\beta$,
the branching ratio is to a good approximation proportional to $\tan^2\beta$, as we can see from the plots. Also this feature is common to all models.
 
\begin{figure}[h!]
 \centering
  \subfigure[$\tan\beta=2$ and $m_0=200\GeV$]
   {\includegraphics[width=7.5cm]{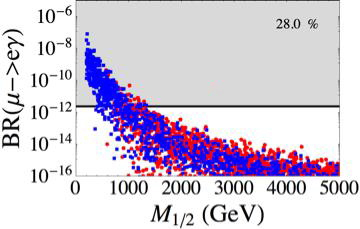}}
  \subfigure[$\tan\beta=2$ and $m_0=5000\GeV$]
   {\includegraphics[width=7.5cm]{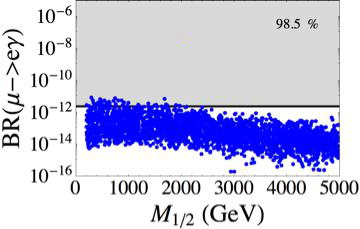}}
  \subfigure[$\tan\beta=15$ and $m_0=200\GeV$]
   {\includegraphics[width=7.5cm]{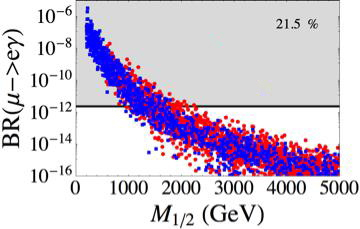}}
  \subfigure[$\tan\beta=15$ and $m_0=5000\GeV$]
   {\includegraphics[width=7.5cm]{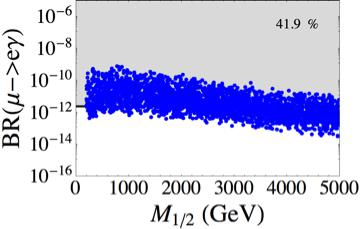}}
 \caption{{\bf Special {\boldmath$A_4$\unboldmath} Models.} Scatter plots of $BR(\mu\to e \gamma)$ as a function of $M_{1/2}$, for different values of $\tan\beta$, and $m_0$. The parameter $|\xi'|$ is chosen as $0.184$ in order to maximize the success rate of this model. The horizontal line shows the current MEG bound. For Blue (Red) points the LSP is the lightest neutralino (stau). The percentage in each plot refers to the number of Blue points that satisfy the MEG bound over the total number of points.}
 \label{fig:MEGSpecialA4Wein}
\end{figure}

For special $A_4$ models the parameter $\xi'$ is taken equal  to $0.184$, as analyzed in Sec.~\ref{sec:SpecialA4}. The results of our numerical study considering the constraint from $BR(\mu\to e\gamma)$ for the $m_0-M_{1/2}$ parameter space are shown in Fig.~\ref{fig:MEGSpecialA4Wein}.  
The plots in Fig.~\ref{fig:MEGSpecialA4Wein} are very similar to those in Fig.~\ref{fig:MEGTypicalA4Wein}. Indeed also for this case, for $\tan\beta=2$ and $m_0=200\GeV$, Fig.~\ref{fig:MEGSpecialA4Wein}(a), almost all the points are excluded for $M_{1/2}\lesssim400\GeV$. The supersymmetric spectrum is also quite similar: the only differences are in the range of masses that the lightest LH and RH charged sleptons can span, $[260,\,500]\GeV$ and $[180,\,340]\GeV$, respectively. By increasing $m_0$ up to $5000\GeV$, Fig.~\ref{fig:MEGTypicalA4Wein}(b), we can see that for the whole range of $M_{1/2}$, the MEG bound is well satisfied in this model, while the corresponding supersymmetric spectrum is heavier.
Special and typical $A_4$ models have a dominant $LL$ mass insertion proportional to $\xi'^2$ and $\xi^2$, respectively, which represents the main difference between
the two models, as far as radiative charged lepton decays are considered. The optimal values of $\xi$ and $|\xi'|$ differ by a factor of about two and we expect that
the branching ratios of the two models should differ by about one order of magnitude, for fixed values of the other parameters. This effect is barely visible in our plots,
due to the spread of the predictions caused by the unknown order-one coefficients.

Increasing the value of $\tan\beta$, the number of points below the MEG bound decreases, but the previous consideration are approximatively still valid: the most interesting difference is in Fig.~\ref{fig:MEGSpecialA4Wein}(c), for $m_0=200\GeV$, where almost all the points with $M_{1/2}<1000\GeV$ are excluded; furthermore, in Fig.~\ref{fig:MEGSpecialA4Wein}(d), for $m_0=5000\GeV$, there are always points satisfying the MEG bound, even if the largest part are excluded, especially for smaller $M_{1/2}$.

\begin{figure}[h!]
 \centering
  \subfigure[$\tan\beta=2$ and $m_0=200\GeV$]
   {\includegraphics[width=7.5cm]{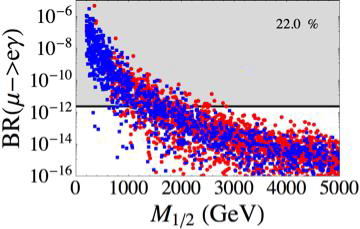}}
  \subfigure[$\tan\beta=2$ and $m_0=5000\GeV$]
   {\includegraphics[width=7.5cm]{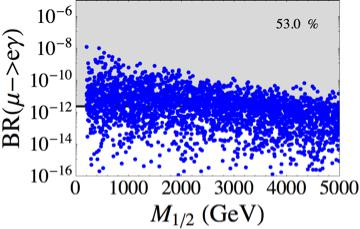}}
  \subfigure[$\tan\beta=15$ and $m_0=200\GeV$]
   {\includegraphics[width=7.5cm]{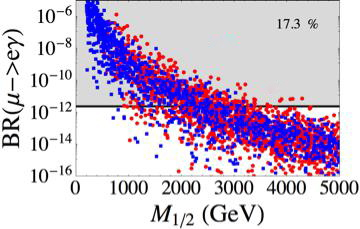}}
  \subfigure[$\tan\beta=15$ and $m_0=5000\GeV$]
   {\includegraphics[width=7.5cm]{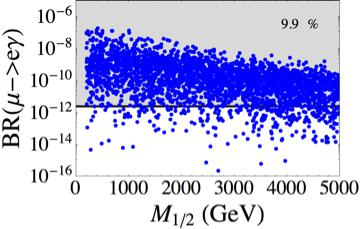}}
 \caption{{\bf {\boldmath$S_4$\unboldmath} Models.} Scatter plots of $BR(\mu\to e \gamma)$ as a function of $M_{1/2}$, for different values of $\tan\beta$, and $m_0$. The parameter $\xi$ is chosen as $0.172$ in order to maximize the success rate of this model. The horizontal line shows the current MEG bound. For Blue (Red) points the LSP is the lightest neutralino (stau). The percentage in each plot refers to the number of Blue points that satisfy the MEG bound over the total number of points.}
 \label{fig:MEGS4BMWein}
\end{figure}

Finally, in the $S_4$ model the parameter $\xi$ is taken equal  to $0.172$, as required to maximize the success rate of these models to arrange the three mixing angles in the corresponding $3\sigma$ ranges. The results are displayed in Fig. \ref{fig:MEGS4BMWein}.
The plots in Fig.~\ref{fig:MEGS4BMWein} are very similar to those in Fig.~\ref{fig:MEGSpecialA4Wein} and the same comments also apply here. More interestingly, in all the plots the number of points satisfying the MEG bound is much smaller, especially for large $m_0$. In particular for $\tan\beta=15$ and $m_0=5000\GeV$, only the $\sim10\%$ of points correspond to a BR smaller than the MEG bound.
The relatively larger branching ratio for $\mu\to e \gamma$ predicted by the $S_4$ model is also a consequence of the scaling of $(\delta_{\mu e})_{LL}$ with respect to $\xi$:
such a scaling is linear in $S4$, while it is quadratic in the typical $A_4$ models. Moreover the optimal value of $\xi$ in $S_4$ is larger than in typical $A_4$ models.
This explains the enhancement by two order of magnitude of the $S_4$ prediction compared to the typical $A_4$ one.

%%%%%%%%%%%%%%%%%%%%%%%%%   3.3 MI from high-energy RGE
\subsection{Mass Insertion from high-energy RGE}

When neutrino masses originate from a Type I See-Saw mechanism, there is an extra contribution to the running of the slepton mass matrices, 
originating from loop diagrams with the exchange of right-handed neutrinos. Such a contribution is only relevant in the energy range from the cut-off $\Lambda$ of the theory 
down to the right-handed neutrino masses $M_k$ $(k=1,2,3)$. By focussing on the LL block $m^2_{eLL}$ of the slepton mass matrix, whose non-diagonal entries dominate
the rates of the processes under consideration, in the leading log approximation we have \cite{Borzumati:1986qx,Gabbiani:1988rb,Petcov:2003zb}
\begin{equation}
\left(m^2_{eLL}\right)_{ij}\simeq-\dfrac{1}{8\pi^2}\left(3\,m_0^2+A_0^2\right)\sum_k(\hat{Y}^\dag_\nu)_{ik} \log\left(\dfrac{\Lambda}{M_k}\right) (\hat{Y}_\nu)_{kj}\,,
\label{m2eLLapproxLL}
\end{equation}
where $A_0$ is the SUSY breaking parameter characterizing the size of the trilinear scalar mass term for sleptons and the matrix $\hat{Y}_\nu$ denotes the neutrino Yukawa couplings in the basis where the mass matrices for charged leptons and right-handed neutrinos have been diagonalized. The above expression for $m^2_{eLL}$
holds at the scale equal to the lightest right-handed neutrino mass. Below that scale, right-handed neutrinos do not affect the running any more.
In the models we are considering neutrino Yukawa couplings are of order one and even for relatively small ratios $\Lambda/M_K\approx 100$, we may easily get
contributions to the LL mass insertions of order 0.1. 

It is interesting to note that if neutrinos transform as an irreducible triplet of the flavour symmetry then, at the LO in the flavour symmetry breaking parameters,
the right-hand side of eq. (\ref{m2eLLapproxLL}) can be expressed in term of the light neutrino masses and the matrix elements of the lepton mixing matrix $U$.
We have
\be
\hat{Y}_\nu=k~ U^\dagger+...
\label{YU}
\ee
and 
 \begin{equation}
\left(m^2_{eLL}\right)_{ij}\simeq-\dfrac{|k|^2}{8\pi^2}\left(3\,m_0^2+A_0^2\right)
\left[ U_{i2} \log\frac{m_2}{m_1} U_{j2}^*+U_{i3}\log\frac{m_3}{m_1} U_{j3}^*
\right]+...\,,
\label{m2eLLirr}
\end{equation}
where $k$ is a constant of order one and dots stand for non-leading contributions, suppressed by powers of $\langle\Phi\rangle/\Lambda$.
This result is completely general. It applies to any model with a flavour symmetry, independently of the specific flavour group $G_f$, provided
neutrinos are assigned to an irreducible triplet of the group, like in the models under consideration.
First note that the invariance of the theory under transformations of $G_f$ implies
\be
\rho(g)~Y^\dagger_\nu Y_\nu~\rho(g)^\dagger=Y^\dagger_\nu Y_\nu\,,
\ee
$\rho$ denoting the irreducible triplet representation under which the left-handed leptons transform.
The combination $Y^\dagger_\nu Y_\nu$ commutes with each group element $\rho(g)$ and, by the Shur's First Lemma, $Y^\dagger_\nu Y_\nu$ either vanishes, a case that we exclude, or is proportional to the unit matrix.
We conclude that $Y_\nu$ is proportional to a unitary matrix. This result holds in any basis, since the invariance of the theory is
a basis-independent property. If now we go to the basis where charged leptons and right-handed neutrinos are mass eigenstates,
the See-Saw relation reads:
\be
m_\nu=\frac{v^2}{2}\hat{Y}_\nu^T M^{-1} \hat{Y}_\nu
\ee 
and we recognize that in this basis the unitary matrix to which $\hat{Y}_\nu$ is proportional should coincide with $U^\dagger$. We obtain eq. (\ref{YU})
and 
\be
M^{-1}=\frac{2}{|k|^2 v^2} (m_\nu)_{diag}
\label{MMO}
\ee
By making use of eqs. (\ref{YU}), (\ref{MMO}) and (\ref{m2eLLapproxLL}) we immediately get the result in eq. (\ref{m2eLLirr}).
Thus, up to sub-leading corrections and up to the unknown order-one parameter $k$, the LL mass insertions are completely determined
by neutrino masses and mixing parameters. At the LO the MIs do not depend on the cut-off scale $\Lambda$, but only on the ratios
between light neutrino masses. For a degenerate neutrino spectrum the LO MIs vanish. On the contrary the largest MIs are obtained
when the spectrum is hierarchical. 

\begin{figure}[h!]
 \centering
  \subfigure[$\tan\beta=2$ and $m_0=200\GeV$]
   {\includegraphics[width=7.5cm]{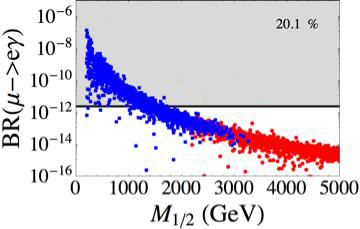}}
  \subfigure[$\tan\beta=2$ and $m_0=5000\GeV$]
   {\includegraphics[width=7.5cm]{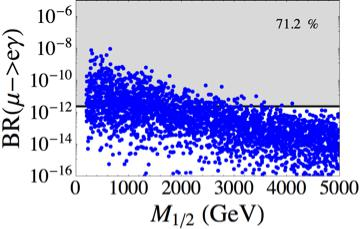}}
  \subfigure[$\tan\beta=15$ and $m_0=200\GeV$]
   {\includegraphics[width=7.5cm]{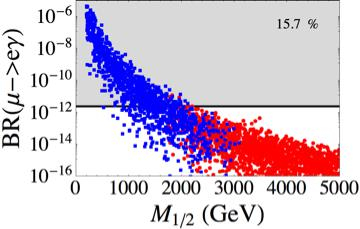}}
  \subfigure[$\tan\beta=15$ and $m_0=5000\GeV$]
   {\includegraphics[width=7.5cm]{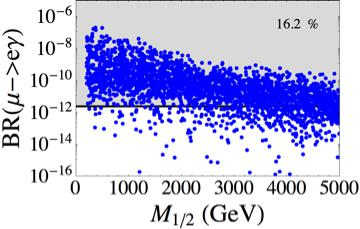}}
 \caption{{\bf Typical {\boldmath$A_4$\unboldmath} Models.} Scatter plots of $BR(\mu\to e \gamma)$ with running effects due to right-handed neutrinos, as a function of $M_{1/2}$, for different values of $\tan\beta$, and $m_0$. 
 Light neutrinos have normal mass ordering witn $m_1=4.4$ meV.
 The parameter $\xi$ is chosen as $0.076$ in order to maximize the success rate of this model. The horizontal line shows the current MEG bound. For Blue (Red) points the LSP is the lightest neutralino (stau). The percentage in each plot refers to the number of Blue points that satisfy the MEG bound over the total number of points.}
 \label{fig:MEGTypicalA4WeinRH}
\end{figure}

We have considered the type I See-Saw version of the previous models and we have evaluated the slepton mass matrices by
including the effects on the running due to right-handed neutrinos. Our plots and our numerical results have been worked out 
at the NLO in the symmetry breaking parameters. RGE equations are solved numerically, making use of full one-loop beta functions. The evolution starts at $\Lambda=2\times 10^{16}$ GeV, where we assume the pattern dictated by the eqs. (\ref{TA4slm1})-(\ref{TA4slm3}), (\ref{SA4slm1})-(\ref{SA4slm3}) and (\ref{S4slm1})-(\ref{S4slm3}) 
with $m_{SUSY}$ identified with $m_0$. Going down to the electroweak scale the off-diagonal entries of the slepton mass matrices are modified by the running due to both 
right-handed neutrinos and by the MSSM degrees of freedom. We display our results for the case of normal ordering.  Similar considerations hold when the neutrino mass ordering is inverted. For the lightest neutrino mass we chose the smallest value allowed by the models under consideration: 
4.4 meV for the typical $A_4$ models, 0.4 meV for the special $A_4$ models and 1 meV for $S_4$. These values can be estimated by analyzing the neutrino masses in the See-Saw version of the models in Refs. 
\cite{Altarelli:2005yx}, \cite{Lin:2009bw} and \cite{Altarelli:2009gn}, considered in the previous subsections. Choosing the smallest value of $m_1$ enhances the ratios $m_2/m_1$ and $m_3/m_1$ in eq. (\ref{m2eLLirr}) and 
maximizes the effect of the running due to right-handed neutrinos.

\begin{figure}[h!]
 \centering
  \subfigure[$\tan\beta=2$ and $m_0=200\GeV$]
   {\includegraphics[width=7.5cm]{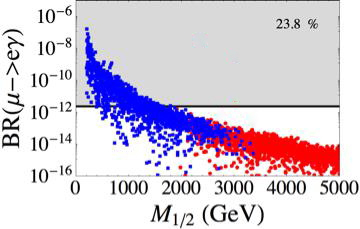}}
  \subfigure[$\tan\beta=2$ and $m_0=5000\GeV$]
   {\includegraphics[width=7.5cm]{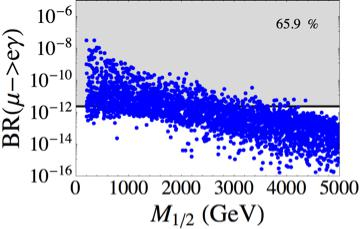}}
  \subfigure[$\tan\beta=15$ and $m_0=200\GeV$]
   {\includegraphics[width=7.5cm]{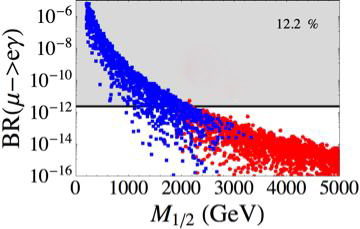}}
  \subfigure[$\tan\beta=15$ and $m_0=5000\GeV$]
   {\includegraphics[width=7.5cm]{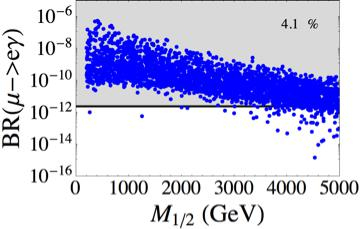}}
 \caption{{\bf Special {\boldmath$A_4$\unboldmath} Models.} 
 Scatter plots of $BR(\mu\to e \gamma)$ with running effects due to right-handed neutrinos, as a function of $M_{1/2}$, for different values of $\tan\beta$, and $m_0$. 
 Light neutrinos have normal mass ordering witn $m_1=0.4$ meV. The parameter $|\xi'|$ is chosen as $0.184$ in order to maximize the success rate of this model. The horizontal line shows the current MEG bound. For Blue (Red) points the LSP is the lightest neutralino (stau). The percentage in each plot refers to the number of Blue points that satisfy the MEG bound over the total number of points.}
 \label{fig:MEGSpecialA4WeinRH}
\end{figure}

\begin{figure}[h!]
 \centering
  \subfigure[$\tan\beta=2$ and $m_0=200\GeV$]
   {\includegraphics[width=7.5cm]{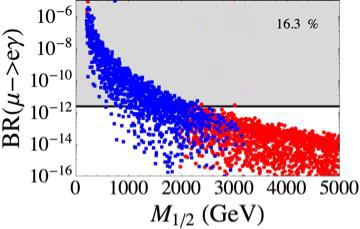}}
  \subfigure[$\tan\beta=2$ and $m_0=5000\GeV$]
   {\includegraphics[width=7.5cm]{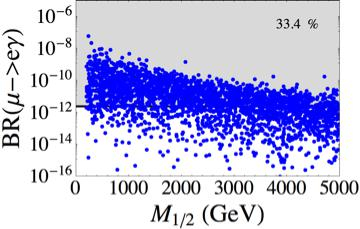}}
  \subfigure[$\tan\beta=15$ and $m_0=200\GeV$]
   {\includegraphics[width=7.5cm]{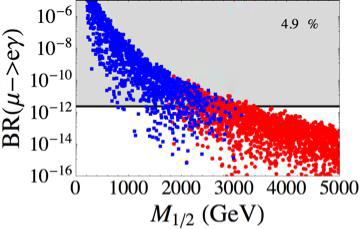}}
  \subfigure[$\tan\beta=15$ and $m_0=5000\GeV$]
   {\includegraphics[width=7.5cm]{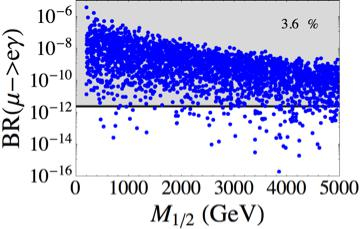}}
 \caption{{\bf {\boldmath$S_4$\unboldmath} Models.} 
 Scatter plots of $BR(\mu\to e \gamma)$ with running effects due to right-handed neutrinos, as a function of $M_{1/2}$, for different values of $\tan\beta$, and $m_0$. 
 Light neutrinos have normal mass ordering witn $m_1=1$ meV.
 The parameter $\xi$ is chosen as $0.172$ in order to maximize the success rate of this model. The horizontal line shows the current MEG bound. For Blue (Red) points the LSP is the lightest neutralino (stau). The percentage in each plot refers to the number of Blue points that satisfy the MEG bound over the total number of points.}
 \label{fig:MEGS4BMWeinRH}
\end{figure}

In Figs.~\ref{fig:MEGTypicalA4WeinRH}, \ref{fig:MEGSpecialA4WeinRH} and \ref{fig:MEGS4BMWeinRH}  we plot the branching ratio for $\mu\to e \gamma$ as a function of $M_{1/2}$ for the same values of $m_0$ and $\tan\beta$ shown in Figs.~\ref{fig:MEGTypicalA4Wein}, \ref{fig:MEGSpecialA4Wein} and \ref{fig:MEGS4BMWein} to allow for a direct comparison between the two cases, with and without right-handed neutrinos. As a general trend, the inclusion of the running due to right-handed neutrinos enhances the branching ratio by a factor between about one and two orders of magnitude. The allowed region in the
parameter space of the models shrinks, as indicated by the percentage of points that satisfy the current MEG bound. The largest effect occurs for the special $A_4$ model, also due to the small value $m_1=0.4$ meV,
which, for normal hierarchy, enhances the contribution in eq. (\ref{m2eLLirr}). In the cases of typical $A_4$ models and $S_4$ models, the effect of right-handed neutrinos is similar. 
Notice that the slope of the plotted regions for $m_0=200$ GeV is much steeper than for $m_0=5000$ GeV which increases the impact of right-handed neutrinos in the latter case. This is particularly visible in the case of special $A_4$ models. If we compare the panel  (d) of Figs.~\ref{fig:MEGSpecialA4Wein} and \ref{fig:MEGSpecialA4WeinRH}, where $m_0=5000$ GeV and $\tan\beta=15$, we see that
the points are lying on an almost horizontal strip close to the MEG bound and the enhancement due to right-handed neutrinos is sufficient to displace almost all the band above the bound, 
thus excluding most of the points. In conclusion, running effects due to right-handed neutrinos can significantly contribute to the off-diagonal terms of slepton mass matrices and can even dominate MIs, especially
for a pronounced hierarchy in the light neutrino mass spectrum. There is a general reduction in the parameter space of the model and in particular for the special $A_4$ model with the most hierarchical spectrum. A milder impact is expected for 
a neutrino mass spectrum close to the degenerate case.

%%%%%%%%%%%%%%%%%%%%%%%%%   3.3 Correlation with the Muon $g-2$
\subsection{Correlation with the Muon $g-2$}

The value found for the anomalous magnetic moment of the muon \cite{Bennett:2006fi}
\beq
a_\mu^{EXP}=116592080(63)\times 10^{-11}
\eeq
shows a 3.4 $\sigma$ deviation 
\beq
\delta a_\mu=a_\mu^{EXP}-a_\mu^{SM}=+302(88)\times 10^{-11}
\label{Da}
\eeq
from the value expected in the SM \cite{Hagiwara:2006jt,Passera:2008jk,Passera:2008hj} 
\beq
a_\mu^{SM}=116591778(61)\times 10^{-11} \; .
\eeq
It is an interesting question whether the presence of SUSY particles can account for this deviation, once the constrains from the branching ratio of the $\mu\to e	\gamma$ decay are taken into consideration.

Following Refs.~\cite{Moroi:1995yh,Martin:2001st,Stockinger:2006zn,Czarnecki:2001pv}, we study the correlation between $\delta a_\mu$ and $BR(\mu\to e\gamma)$, for $\tan\beta\in[2,\,15]$ and $m_0,\,M_{1/2}\in[200,\,5000]\GeV$, while all the other parameters are treated according to the previous section. Only points corresponding to scenarios with the lightest neutralino being the LSP are shown.

\begin{figure}[h!]
 \centering
 \subfigure[Typical $A_4$. $\xi=0.076$.]
   {\includegraphics[width=5.3cm]{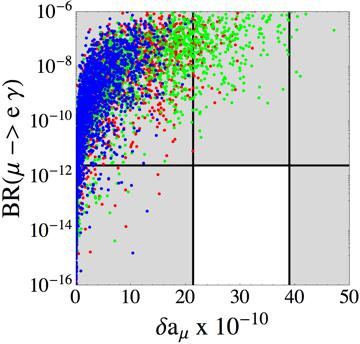}}
 \subfigure[Special $A_4$. $|\xi'|=0.184$.]
   {\includegraphics[width=5.3cm]{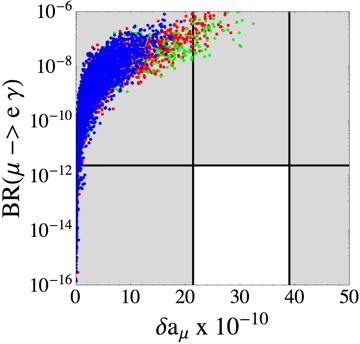}}
 \subfigure[$S_4$. $\xi=0.172$.]
   {\includegraphics[width=5.3cm]{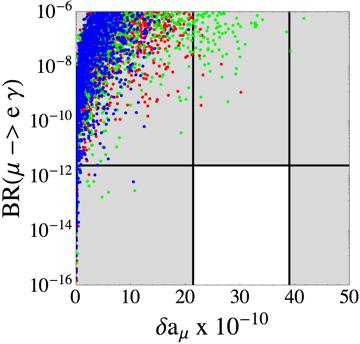}}
\caption{Correlation plots between $\delta a_\mu$ and $BR(\mu\to e\gamma)$. The value of $\tan\beta$ is taken in the range $[2,\,15]$, while $m_0,\,M_{1/2}$ are chosen between $200$ and $5000$ GeV. The values of $\xi$ or $|\xi'|$ are those that maximize the success rates for the three models for the NH case. The colour of the points refer to the values of $\tan\beta$: $2\lesssim\tan\beta\lesssim7$ in blue, $7\lesssim\tan\beta\lesssim11$ in red, $11\lesssim\tan\beta\lesssim15$ in green. The horizontal line corresponds to the MEG bound, while the vertical lines correspond to the measurements on $\delta a_\mu$ at $3\sigma$.}
\label{ScatterBrG2}
\end{figure}

As we can see from Fig.~\ref{ScatterBrG2}, in the whole parameter space of the three models it is not natural to reproduce the observed deviation of the muon anomalous magnetic moment, once we consider the present MEG bound on $BR(\mu\to e \gamma)$. This is not a surprise, because the explanation of the $3.4 \sigma$ discrepancy needs small values of $m_0$ and $M_{1/2}$ and larger
values of $\tan\beta$, which, however, enhance the branching ratio of the radiative LFV decays.

%%%%%%%%%%%%%%%%%%%%%%%%%%%%%%%%%%%%%%%%%%%%%%%%%%%%%%%%%%%%%%%%%
%%%%%%%%%%%%%%%%%%%%%%%%%   5. Conclusion     
%%%%%%%%%%%%%%%%%%%%%%%%%%%%%%%%%%%%%%%%%%%%%%%%%%%%%%%%%%%%%%%%%
\section{Conclusion}
\label{sec:Conclusions}

The recent rather precise measurements of $\theta_{13}$ make our present knowledge of the neutrino mixing matrix, except for the CP violating phases, sufficiently complete to considerably restrict the class of models that can reproduce the data. In spite of this progress, the range of possibilities for flavour models remains unfortunately quite wide. On the one extreme, the rather large value measured for $\theta_{13}$, close to the old CHOOZ bound, has validated the prediction of models based on anarchy \cite{Hall:1999sn,deGouvea:2003xe}, i.e. no symmetry in the leptonic sector, only chance, so that this possibility remains valid, as discussed, for example, in Ref. \cite{deGouvea:2012ac}.  Anarchy can be formulated in a $SU(5) \otimes U(1)$ context by taking different Froggatt-Nielsen\cite{Froggatt:1978nt} charges only for the $SU(5)$ tenplets (for example $10\sim(3,2,0)$, where 3 is the charge of the first generation, 2 of the second, zero of the third) while no charge differences appear in the $\bar 5$ ($\bar 5\sim (0,0,0)$). Anarchy can be mitigated by assuming that it  only holds in the 2-3 sector with the advantage that the first generation masses and the angle $\theta_{13}$ are naturally small (see also the recent revisiting in Ref.\cite{Buchmuller:2011tm}). In models with See-Saw, one can also play with the charges for the right-handed SU(5) singlet neutrinos. If, for example, one takes $1\sim(1, -1, 0)$, together with $\bar 5\sim (2,0,0)$,  it is possible to get a normal hierarchy model with $\theta_{13}$ small and also with $r = \Delta m^2_{solar}/\Delta m^2_{atm}$ naturally small (see, for example, Ref.~\cite{Altarelli:2002sg}). In summary, anarchy and its variants, all based on chance, offer a rather economical class of models that are among those encouraged by the new $\theta_{13}$ result. On the other extreme, stimulated by the fact that the data suggest some special mixing patterns as good first approximations (TB or BM, for example), models based on discrete flavour symmetries, like $A_4$ or $S_4$, have been proposed and widely studied. In these models the starting LO approximation is completely fixed (no chance), but the NLO corrections introduce a number of undetermined parameters. The recent data on $\theta_{13}$ and the MEG new upper bound on the LFV process $\mu \to e \gamma$ impose a reappraisal of these models, which we have attempted in this paper. In particular, the relatively large value of $\theta_{13}$ introduces a marked departure from the TB limit, while the values of $\theta_{12}$ and $\theta_{23}$ are very close to it. The challenge is to produce in a natural way a relatively large correction to $\theta_{13}$ without affecting too much the other mixing angles. But one must pay attention that these larger corrective terms introduced to shift $\theta_{13}$ from the TB value could appear in the non-diagonal elements of the charged lepton (and s-lepton) mass matrix and could induce a too large $\mu \to e \gamma$ branching ratio. 

As a result of our analysis we find that, for reproducing the mixing angles, the Lin type $A_4$ models have the best performance, as expected, followed by the typical $A_4$ models, while the BM mixing models lead to an inferior score, as they most often fail to reproduce $\theta_{12}$. In the latter case, if only one complex parameter perturbs the BM pattern, there is a strict correlation among the solar and the reactor angle and the success rate increases considerably by selecting a CP violating Dirac phase close to $\pi$.

As for LFV processes we have addressed the problem by adopting the simple CMSSM framework. While this overconstrained version of supersymmetry is rather marginal after the results of the LHC searches, more so if the Higgs mass really is around $m_H=125$ GeV, we still believe it can be used here for our indicative purposes. We find that the most constrained versions are the models that start with BM mixing at the LO because, in this case, relatively large corrections directly appear in the off-diagonal terms of the charged lepton mass matrix. The typical $A_4$ models turn out to be the best suited to satisfy the MEG experimental bound, as the non-diagonal charged lepton matrix elements needed to reproduce the mixing angles are quite smaller. An intermediate, still rather good, score is achieved by the models of the Lin type, where the main corrections to the mixing angles arise from the neutrino sector. When the fit to the mixing angles and the bounds on LFV processes are combined, the $A_4$ models emerge well from our analysis and in particular those of the Lin type perhaps appear as the most realistic approach to the data among the models based on discrete flavour groups that we have studied.  As for the regions of the CMSSM parameter space that are indicated by our analysis the preference is for small $\tan{\beta}$ and large SUSY masses (at least one out of $m_0$ and $m_{1/2}$ must be above 1 TeV).  As a consequence it appears impossible, at least within the CMSSM model, to satisfy the MEG bound and, at the same time, to reproduce the muon $g-2$ discrepancy.

%%%%%%%%%%%%%%%%%%%%%%%%%%%%%%%%%%%%%%%%%%%%%%%%%%%%%%%%%%%%%%%%
% Acknowledgements
%%%%%%%%%%%%%%%%%%%%%%%%%%%%%%%%%%%%%%%%%%%%%%%%%%%%%%%%%%%%%%%%
\section*{Acknowledgements}

We recognize that this work has been partly supported by the Italian Ministero dell'Uni\-ver\-si\-t\`a e della Ricerca Scientifica, under the COFIN program (PRIN 2008), by the European Commission, under the networks ``Heptools'', ``Quest for Unification'', ``LHCPHENONET'' and European Union FP7  ITN INVISIBLES (Marie Curie Actions, PITN- GA-2011- 289442) and contracts MRTN-CT-2006-035505 and  PITN-GA-2009-237920 (UNILHC), and by the Te\-ch\-ni\-sche Universit\"at M\"unchen -- Institute for Advanced Study, funded by the German Excellence Initiative.

%%%%%%%%%%%%%%%%%%%%%%%%%%%%%%%%%%%%%%%%%%%%%%%%%%%%%%%%%%%%%%%%
% Appendix
%%%%%%%%%%%%%%%%%%%%%%%%%%%%%%%%%%%%%%%%%%%%%%%%%%%%%%%%%%%%%%%%
\appendix
\boldmath
\section{Expression of $a_{CC^\prime}$}
\unboldmath

In this section we list the explicit expression of $a_{CC^\prime}$ in our conventions:
\beq
\begin{aligned}
a_{LL}&=\dd\frac{g^2}{16\pi^2}\left[ f_{1n}(a_2)+f_{1c}(a_2)+
\dd\frac{M_2\mu\tan\beta}{M_2^2-\mu^2}\Big(f_{2n}(a_2,b)+f_{2c}(a_2,b)\Big)\right.\\
&+\left.\tan^2\theta_W\left(f_{1n}(a_1)- \dd\frac{M_1\mu\tan\beta}{M_1^2-\mu^2}f_{2n}(a_1,b)- M_1\left(m_{SUSY}-\mu\tan\beta\right)\dfrac{f_{3n}(a_1)}{m_{SUSY}^2}\right)\right]\\
a_{RL}&=\dd\frac{g^2}{16\pi^2}\tan^2\theta_W\dd\frac{M_1}{m_{SUSY}} 2 f_{2n}(a_1)\\
a_{RR}&=\dd\frac{g^2}{16\pi^2}\tan^2\theta_W \left[4 f_{1n}(a_1)+ 2\dd\frac{M_1\mu\tan\beta}{M_1^2-\mu^2}f_{2n}(a_1,b)- M_1\left( m_{SUSY}-\mu\tan\beta\right)\dfrac{f_{3n}(a_1)}{m_{SUSY}^2}\right]\\
a_{LR}&=\dd\frac{g^2}{16\pi^2}\tan^2\theta_W\dd\frac{M_1}{m_{SUSY}} 2 f_{2n}(a_1)
\end{aligned}
\label{MIcoefficients}
\eeq
where $a_{1,2}=M^2_{1,2}/m_{SUSY}^2$, $b=\mu^2/m_{SUSY}^2$ and $f_{i(c,n)}(x,y)=f_{i(c,n)}(x)-f_{i(c,n)}(y)$.
The functions $f_{in}(x)$ and $f_{ic}(x)$, slightly different from those in Ref.~\cite{Ciuchini:2007ha}, are given by:
\beq
\begin{aligned}
f_{1n}(x)&=(-17 x^3+9 x^2+9 x-1+6 x^2(x+3) \log x)/(24(1-x)^5)\\
f_{2n}(x)&=(-5 x^2+4 x+1+2x(x+2)\log x)/(4(1-x)^4)\\
f_{3n}(x)&=(1+9x -9x^2-x^3+6x(x+1) \log x)/(2(1-x)^5)\\
f_{1c}(x)&=(-x^3-9x^2+9x+1+6x(x+1) \log x)/(6(1-x)^5)\\
f_{2c}(x)&=(-x^2-4 x+5+2(2x+1)\log x)/(2(1-x)^4)\,.
\end{aligned}
\label{MIfunctions}
\eeq

%%%%%%%%%%%%%%%%%%%%%%%%%%%%%%%%%%%%%%%%%%%%%%%%%%%%%%%%%%%%%%%%%
%%%%%%%%%%%%%%%%%%%%%%%%%  Bibliography     
%%%%%%%%%%%%%%%%%%%%%%%%%%%%%%%%%%%%%%%%%%%%%%%%%%%%%%%%%%%%%%%%%

%\bibliography{biblio}{}
%\bibliographystyle{BiblioStyle}

\providecommand{\href}[2]{#2}\begingroup\raggedright\endgroup

\end{document}